\documentclass[a4paper,oneside]{scrartcl}

\usepackage[english]{babel}
\usepackage{listings}
\usepackage{xcolor}
\usepackage{booktabs}
\usepackage{graphicx}
\usepackage{url}
\usepackage{hyperref}
\usepackage{authblk}
\usepackage{abstract}
\usepackage[numbers]{natbib}
\usepackage{verbatim}
\usepackage{pifont}
\usepackage{lipsum}
\usepackage[caption=false]{subfig}
\usepackage{xspace}
\usepackage{calc}

\begin{document}

\title{\mbox{}LLAMA: The Low-Level Abstraction for Memory Access}

\author[1,2,3]{Bernhard Manfred Gruber} 
\author[1]{Guilherme Amadio} 
\author[1]{Jakob Blomer} 
\author[4,5]{Alexander Matthes} 
\author[4]{René Widera} 
\author[2]{Michael Bussmann} 

\date{2. February 2022}

\affil[1]{EP-SFT, CERN, Geneva, Switzerland}
\affil[2]{Center for Advanced Systems Understanding (CASUS), Saxony, Germany}
\affil[3]{Faculty of Computer Science, Technische Universität Dresden, Dresden, Germany}
\affil[4]{Helmholtz-Zentrum Dresden - Rossendorf (HZDR), Dresden, Germany}
\affil[5]{LogMeIn, Dresden, Germany}

\maketitle

\begin{abstract}
	The performance gap between CPU and memory widens continuously.
	Choosing the best memory layout for each hardware architecture is increasingly important as more and more programs become memory bound.
	For portable codes that run across heterogeneous hardware architectures, the choice of the memory layout for data structures is ideally decoupled from the rest of a program.
	This can be accomplished via a zero-runtime-overhead abstraction layer, underneath which memory layouts can be freely exchanged.

	We present the low-level abstraction of memory access (LLAMA), a C++ library that provides such a data structure abstraction layer with example implementations for multidimensional arrays of nested, structured data.
	LLAMA provides fully C++ compliant methods for defining and switching custom memory layouts for user-defined data types.
	The library is extensible with third-party allocators.

	Providing two close-to-life examples, we show that the LLAMA-generated AoS and SoA layouts produce identical code with the same performance characteristics as manually written data structures.
	Integrations into the SPEC CPU\textsuperscript{\textregistered} lbm benchmark and the particle-in-cell simulation PIConGPU demonstrate LLAMA's abilities in real-world applications.
	LLAMA's layout-aware copy routines can significantly speed up transfer and reshuffling of data between layouts compared with naive element-wise copying.

	LLAMA provides a novel tool for the development of high-performance C++ applications in a heterogeneous environment.
\end{abstract}

\footnotetext{
	\textbf{Abbreviations:}
	\begin{description}
		\item[AoS] array of structs
		\item[AoSoA] array of structs of arrays
		\item[HBM] high bandwidth memory
		\item[HEP] high energy physics
		\item[HPC] high performance computing
		\item[LLAMA] low‐level abstraction of memory access
		\item[LOCs] lines of code
		\item[PGAS] partitioned global address space
		\item[SCM] storage class memory
		\item[SoA] struct of arrays
	\end{description}
}

\section{Introduction}

Application performance is decreasingly limited by available compute resources.
Many programs nowadays are memory-bound~\cite{reflections_memory_wall,processor_memory_gap,processor_memory_bottleneck}, mainly due to current choices in hardware design.
The efficiency of parallel programs thus increasingly relies on maximizing data throughput by increasing data locality and choosing data layouts for efficient parallel access.

Moreover, the increasing heterogeneity of computing hardware means that optimal access patterns vary between platforms and portable efficient data transport becomes a challenge.
We observe a proliferation of specialized compute and memory hardware architectures to fight the increasing gap between compute and memory capabilities, including Google's Tensor Processing Units~\cite{tpu_perf_analysis}, the PEZY SC many-core microprocessors~\cite{pezy_sc}, Fujitsu's A64FX processor powering the Fugaku supercomputer~\cite{a64fx}, near-memory~\cite{near_memory_computing_archs_review} and in-memory~\cite{in_memory_computing_archs_review} computing architectures,
DRAM modules with additional SRAM caches for their row buffers~\cite{row_buffer_cache},
and others ~\cite{tearing_down_memory_wall_hardware}.

Modern hardware is only used efficiently if its internal structure is respected.
That concerns memory accesses as well as the parallel execution of programs.
In modern hardware architectures we must consider cache hierarchies, latencies, bandwidths, volatile memory technology (e.g. DDR, GDDR, HBM), byte-addressable non-volatile memory architectures (e.g. SCM), programmable scratchpad memories (e.g. CUDA shared memory), memory sizes (e.g. L1 cache size) and the type of access (e.g. read, write, atomic, uncached, streaming, non-temporal).
Spatial and temporal access patterns of the algorithm need to be considered as well.

This presents a problem if a portable and performant program is desired: A chosen data structure can only be optimized for one hardware architecture.
For specific domains, small to medium-size applications and a restricted set of platforms, the needed set of memory layouts can still be implemented and optimized, requiring separate code paths depending on the target system, negatively impacting maintainability.
By contrast, general-purpose and broadly applicable data structures entail sub-optimal performance for specific use cases or platforms, and frequently lack flexibility and tunability on the user side.

The efficient copying and reshuffling of data represents an additional major challenge, especially in a heterogeneous memory hierarchy, where different memory layouts of the same data will be used.
Inter-device copies benefit from transferring data in larger blocks, while optimal memory layouts will likely differ on very fine levels.

Maximizing data locality, optimizing data access and data movement for throughput critically depends on having full control over data layouts.
A successful approach retains the expressive strength of general-purpose programming facilities for describing and accessing data structures, and at the same time allows for efficient memory access of these.
In many mainstream languages though, the memory layout cannot be separated from a data structure.
New, possibly domain specific languages or code generative approaches could solve this issue, but often lock themselves out of established ecosystems dominated by traditional languages and frameworks written in them.
A solution should be widely available on a broad range of systems, using and integrating with established and standard technologies.
Thus, we chose C++ as implementation and target language for such a solution due to C++'s wide-spread usage in high performance computing and data analysis.
C++ also provides the necessary language facilities to allow a high-level abstraction at a high execution speed with little to zero overhead on a broad range of hardware platforms, including CPUs, GPUs and FPGAs.

A C++ based solution needs to allow users to express a data structure in a close-to-native way and independently of how it is stored.
Consequently, programs written against this data structure's interface are not bound to the data structure's layout in memory.
This requires a data layout independent way to access the data structure.
Data access needs to be potentially lazily evaluated to defer the target address computation and the actual data fetch as much as possible, allowing the solution to have a bigger picture of the performed memory access and only touch memory regions which are actually needed.

The solution needs to provide generic facilities to map user-defined data structures to performant data layouts while also allowing specialization of this mapping for specific data structures.
A data structure's mapping must be chosen and resolved statically at compile time to give the compiler full optimization potential, thus guaranteeing the same performance as manually written versions of a data structure.

The solution should identify the set of information on hardware structure and algorithmic memory access required for optimizations.
This meta information should be provided as input to the memory mappings, enabling to choose and specialize an optimal memory mapping for a given hardware architecture automatically.

Once a solution has the ability to fully describe a memory mapping in a consumable way, and meta information on hardware and program access pattern is provided, the path is open to provide efficient, high-throughput copying between different data layouts.
This includes deep copies, zero copies and in-situ transformation of data layouts.
A good solution for heterogeneous environments will identify large, contiguous, common chunks of memory between different memory layouts to reduce the number of necessary copy commands for inter-device copies.

It is our firm belief that optimizing access and throughput on the available systems will focus on optimizing the structure of small data blocks first, targeting the fastest memory with highest throughput and lowest latency.
Data layouts are then built bottom-up to respect and map optimally to a target hierarchy of memory systems.
A synthesis of these data layouts into more complex and perhaps distributed data structures and assuring for a global optimization requires different solutions, which are outside the scope of this article.

The implementation of such a solution in a close-to-native, familiar and usable way is challenging, as it breaks C++'s paradigm that a data structure definition is strictly tied to a memory layout by the language's rules.

A well-designed solution will handle memory layout separately from allocation, thus enabling integration with user-provided allocators and  allowing the solution's use in mobile, embedded and freestanding environments, where dynamic memory allocation is often prohibited.

Our contribution, the low-level abstraction of memory access (LLAMA), is an attempt to provide such a solution as a C++ library.
LLAMA grew out of the needs and technologies developed within PIConGPU, a highly performant particle-in-cell simulation framework for CPUs, GPUs and many-core architectures~\cite{picongpu}.

\section{Related work}

Several attempts have been made to tackle memory layout abstraction and portable optimization, often tied to specific usage scenarios.
In the following we provide an overview of relevant categories of libraries and technologies and point to some prominent examples.
This overview is not meant to be exhaustive, but rather give a starting point for the interested reader.
Table \ref{tbl:feature_comparison} summarizes the features of these libraries and technologies discussed in this section.

We firmly believe that it is crucial to allow a flexible, problem-specific definition of data structures on the user side, while allowing for optimization of data access and throughput on the hardware side.
This will become increasingly relevant with growing complexity of simulation and data analysis code, and also when democratizing access to HPC, allowing domain experts without strong HPC backgrounds to still produce efficient codes.
Many current solutions approach memory layout optimization from the hardware perspective rather than from the user side.
We thus think that providing users with more freedom, expressiveness and flexibility when defining their data structures, while still leaving room for optimization, is a good way forward to broaden the user and application base of HPC resources.

\subsection{SoA containers}
SoA containers provide container class templates which support rearranging the container's value type from the traditional AoS storage (e.g. \lstinline{std::vector<T>}) into a SoA storage.
Such libraries can be found in research~\cite{UltimateSoA, ultimate_container, SOAContainer, soax, data_layout_and_SIMD_abstraction, aos_soa_gpu_gems} as well as in industry~\cite{SDLT}.
These libraries typically focus on CPU targets and automatic vectorization, providing strong performance gains over traditional AoS data structures on linearly-iterating access and especially if only subparts of the inner structure are needed.
If the access is at irregular array positions and to almost all of the inner structure, AoS layouts provide better locality of reference.
SoA libraries typically split the data into multiple memory allocations, which has significant advantages when only parts of the data are needed for bulk processing or IO.
Providing an AoSoA container is harder to implement while retaining successful automatic vectorization~\cite{data_layout_and_SIMD_abstraction}.
Our experience confirms this.
Still, AoSoA layouts address an important sweet spot between SoA and AoS, allowing vectorized/coalesced access but maintain some degree of locality.

LLAMA goes beyond such libraries by allowing arbitrary memory layouts and supporting them on CPUs and accelerators.
The popular AoS, SoA and AoSoA are the most common choices for laying out sequences of structures for CPUs and therefore have direct support in LLAMA.

\subsection{SIMD libraries and AoSoA}

There is an assortment of SIMD libraries~\cite{xsimd, boost_simd, nsimd, eve, vc} available as well as efforts to standardize SIMD types in C++~\cite{parallelism_ts_2}.
These primarily offer abstractions for SIMD vectors, masks, arithmetic operations and mathematical functions, which are resolved to a SIMD technology at compile time.
However, the SIMD vector types are also especially useful to build SIMD compatible data structures.
By putting, possibly heterogeneous, vector types instead of scalars into structures, AoSoA memory layouts are easily defined.
Fetching SIMD vectors from pointers to contiguous data also allows easy interaction with SoA layouts.

The Vc~\cite{vc} library for example provides the explicit \lstinline{simdize<S>} construct to create SIMD versions of existing class templates \lstinline{S<A>} on an arithmetic type \lstinline{A}, for instance turning a struct containing float scalars into a struct containing float vectors.

While AoSoA and SoA data layouts in conjunction with SIMD data types are highly relevant for performant CPU focused data structures, SIMD libraries, which commonly wrap CPU specific vector intrinsics, usually fail to support higher dimensional arrays and accelerators such as GPUs.
LLAMA supports AoSoA and SoA memory layouts of heterogeneous structures as well, but agnostic of any SIMD technology or target hardware architecture and is thus fully portable.

\subsection{Multidimensional array libraries}

There is a class of libraries explicitly focusing on generic multidimensional arrays~\cite{boost_multiarray, array_fire, mdarray}, which are often needed to model scientific problems.
The central design aspect is usually the specification of a storage order, which determines in which order the multidimensional array's indices are linearized into memory, which has profound implications on the memory access pattern.
For 2D arrays, this is well known as row- or column-major layout, but the concept generalizes to higher dimensions.
Some libraries even provide more advanced layouts using space filling curves such as Morton codes or Hilbert curves~\cite{MTL4}, or even support the specification of arbitrary array linearization schemes\cite{mdspan}.

A variant of such multidimensional array libraries are non-owing multidimensional array wrappers, such as \lstinline{std::mdspan}~\cite{mdspan, mdspan_case_study}, which provide a multidimensional array interface on top of existing, flat storage.
This separates array functionality from memory allocation, enabling important use cases where data access needs to be mapped to existing regions of storage handled outside the library, e.g. communication buffers prepared by third-party APIs.

Multidimensional array libraries are typically constructed for scalar value types and offer no functionality to split, possibly nested, structured value types and map parts of them differently.
Also, these libraries resolve to a memory location of their value type immediately on access, before further navigation on the value type can continue.

LLAMA can postpone eager evaluation of an access to a memory location and return an intermediary proxy object, thus avoiding to form memory addresses to full intermediate objects.
This lazy evaluation of access allows richer possibilities for memory layouts like splitting and rearranging the elements of a data structure.
Storage order and space filling curves are also supported as part of LLAMA's mappings.

\subsection{Compute kernel abstraction libraries}

Recent technologies for expressing parallel and portable compute kernels, like Kokkos~\cite{kokkos}, RAJA~\cite{raja} or SYCL~\cite{sycl2020}, need to address portable memory handling as part of their design.

Kokkos provides views as memory and buffer abstraction with a rich set of specializations allowing a good decoupling of algorithm and underlying data storage.
These include views backed by host, device or unified virtual memory, PGAS views using MPI and views of SIMD types.
For views with two or more dimensions, a column/row major, strided or fully user defined layout can be chosen.

RAJA's views are similar to \lstinline{std::mdspan} as they are non-owning and based on existing arrays.
They also have a customizable layout trait, determining how a view maps indices to locations in the underlying arrays, including permuting and offsetting indices and creating subviews with offset indices.
RAJA's multi-views also allow to map the data into multiple allocations.

SYCL employs accessors as abstractions to access buffer and image data.
While supporting multidimensional access, the storage order is fixed to row-major for buffer and to column-major for image accessors.

While portable across more acceleration technologies, Kokkos' and RAJA's views and SYCL suffer from the same limitation as multidimensional array libraries.
SYCL additionally does not allow to customize how multidimensional array indices are linearized.

With the flexibility of LLAMA, there is no longer a need to couple a solution for performance portable memory layout optimization to a specific parallel programming library, and libraries such as Alpaka~\cite{alpaka}, that do not make assumptions on data layout, can be used together with LLAMA.

\subsection{Ecosystems for parallelization of scientific algorithms}

More and more, ecosystems for parallelization of scientific codes come about that provide their own, flexible data layouts for a broad range of applications.
As an example, OpenFPM~\cite{openfpm} provides a scalable and open C++ framework for particle and mesh simulation.
It contains, among many domain specific features, a powerful meta-programming mechanism to define distributed, multidimensional arrays of structured data, which is usable separately from the rest of OpenFPM.
The memory layout, AoS or SoA, can be independently configured by the user via an additional template argument.
The SoA layout is limited to the first nesting level of structured data.
Arbitrary linearization of multidimensional arrays is fully configurable.
CUDA is supported.
OpenFPM also includes carefully tuned copy routines between data structures which are aware of the involved memory layouts.
After consulting with OpenFPM's developers, a generalization to allow any user-defined memory layout would be possible with moderate effort on their side.
Alternatively, the user can also change individual memory layouting aspects via template specialization.

Although OpenFPM provides overlapping functionality in its low-level single-core layer, the project's focus is different.
Whereas OpenFPM centers more around sparse arrays and scalable, distributed data structures with top-down domain decomposition,
LLAMA aims for a much smaller, focused, low-level and bottom-up framework, targeting shared memory machines.
This allows easier and broader usage of the library.
We see LLAMA as a generalization of the memory layout capabilities introduced in OpenFPM's foundation.
A similar case can be made for Cabana~\cite{cabana}, a performance portable library for particle-based simulations, which additionally supports an AoSoA layout.

\begin{center}
	\begin{table}[h]
		\centering
		\definecolor{darkgreen}{RGB}{0,150,0}
		\definecolor{darkred}{RGB}{200,0,0}
		\newcommand{\yes}{{\color{darkgreen}\ding{51}}}
		\newcommand{\no}{{\color{darkred}\ding{55}}}
		\newcommand{\may}{{\color{darkgreen}(\ding{51})}}
		\caption{Comparison of memory layouting capabilities of various C++ libraries and frameworks. \yes denotes supported features, \may = partially supported features and \no marks unsupported features.\label{tab_comparison}}
		\newcommand{\thd}[1]{\textbf{#1}}
		\resizebox{\textwidth}{!}{
		\begin{tabular}{c c c c c c c}
			\toprule
			\thd{Feature} & \thd{SoA cont.} & \thd{Vc SIMD lib} & \thd{MD arrays} & \thd{Kernel abs.} & \thd{OpenFPM} & \thd{LLAMA} \\
			\midrule
			SoA               & \yes & \yes & \no  & \no  & \may & \yes \\
			AoSoA             & \no  & \yes & \no  & \no  & \no  & \may \\
			Multi-dim. arrays       & \no  & \no  & \yes & \yes & \yes & \yes \\
			Any array linearization & \no  & \no  & \yes & \yes & \yes  & \yes \\
			Any user-def. mapping             & \no  & \no  & \no  & \yes & \may  & \yes \\
			(Nested) structures     & \yes & \yes & \no  & \no  & \yes & \yes \\
			Accelerator supp. & \may & \no  & \may & \yes & \yes & \yes \\
			Multiple allocations    & \yes & \no  & \no  & \may & \yes & \yes \\
			Lazy access evaluation  & \may & \no  & \no  & \no  & \may  & \yes \\
			Layout aware copy & \no  & \no  & \no  & \no  & \yes & \yes \\
			Allocator indep. use    & \no  & \yes & \may & \yes & \yes & \yes \\
			\bottomrule
		\end{tabular}
		}
		\let\thd\undefined
		\let\yes\undefined
		\let\no \undefined
		\let\may\undefined
		\label{tbl:feature_comparison}
	\end{table}
\end{center}

\section{Design}

The design of LLAMA centers on an abstract description of a data structure containing compile time and runtime information, which is instantiated with a separately configurable memory layout.
The following sections will elaborate on the conceptual design of LLAMA, introduce its library components and walk the reader through these components in the order they will be used when writing a program with LLAMA.

\subsection{Concept}

\begin{figure}
	\centerline{\includegraphics{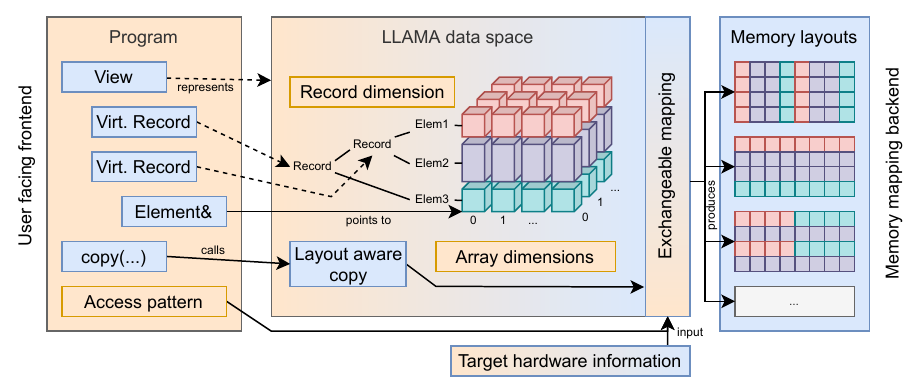}}
	\caption{
		Conceptual overview of LLAMA.
		\label{fig:concept}
	}
\end{figure}

Figure~\ref{fig:concept} illustrates the conceptual design of LLAMA.
In order to separate data structure access and physical memory layout LLAMA introduces an abstract data type called \emph{data space}.
The data space is an index set of hypercubic shape described by the \emph{record dimension} and one or more \emph{array dimensions}.
The record dimension consists of a hierarchy of names and describes nested, structured data, much like a \lstinline{struct} in C++.
LLAMA is thus currently limited to n-dimensional, dynamically-sized arrays of structures.
Programs are written against this abstract data space and thus formulated independently of the physical manifestation of the data space.
Programs can refer to the data space via a \emph{view}, and to subparts of the data space via \emph{virtual records}, which behave like references to \lstinline{struct}s, and to individual elements via real l-value references.
To allow a flexible specification of a memory layout, the data space is materialized via an exchangeable \emph{mapping} that describes how the index set of the data space is embedded into a physical memory.
For further optimization, this mapping can be augmented with additional information from the program's access pattern and target hardware information.
Due to a mapping encapsulating the full knowledge of a memory layout, LLAMA supports \emph{layout-aware copy} operations between instances of the same data space but with different mappings.

This multidimensional index approach mirrors sets of parallel tasks or threads identified by similar multidimensional index structures, as common in e.g. CUDA, OpenCL or Alpaka.

\subsection{Library overview}

\begin{figure}
	\centerline{\includegraphics{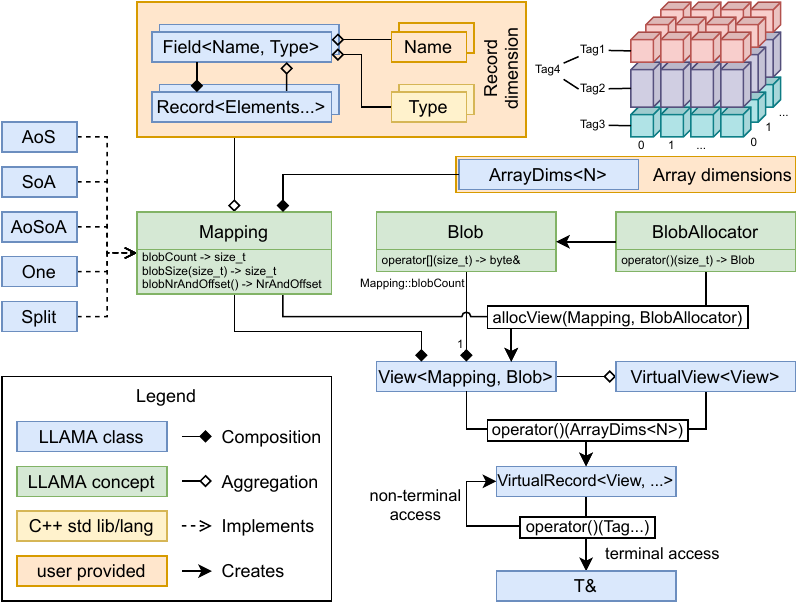}}
	\caption{
		Overview of the C++ library components of LLAMA.
		\label{fig:overview}
	}
\end{figure}

Figure~\ref{fig:overview} gives an overview of the components of LLAMA.
The core data structure of LLAMA is the \lstinline{View}, which provides methods to access the data space.
In order to create a view, a \lstinline{Mapping} is needed.
LLAMA offers many kinds of mappings and users can also provide their own mappings.
Mappings are constructed from a record dimension, containing nested records with named fields holding the element types, and array dimensions.
In addition to a mapping defining the memory layout, an array of blobs is needed for a view, supplying the actual storage behind the view.
A \lstinline{Blob} is any object representing a contiguous chunk of memory.
A suitable \lstinline{Blob} array is either directly provided by the user or built using a \lstinline{BlobAllocator}.
This allows LLAMA to stay independent of and orthogonal to allocators.

The user can navigate and drill down on their data starting at a view.
Created on top of a \lstinline{View}, a \lstinline{VirtualView} restricts access to a subspace of the array dimensions.
Elements of the array dimensions, called records, are accessed on both, \lstinline{View} and \lstinline{VirtualView}, by specifying an array dimensions coordinate.
This access returns a \lstinline{VirtualRecord}, allowing further access using the tags from the record dimension, until eventually a reference to actual data in memory is returned.

\subsection{Describing a data structure}

LLAMA distinguishes between the array and the record dimensions.
The array dimensions are defined at runtime whereas the record dimension is defined at compile time.
This allows to make the problem size itself a run time value but leaves the compiler room to optimize the data access.

The record dimension is statically sized and the equivalent of nested \lstinline{struct}s in C++.
A \lstinline{Record<Fields...>} defines a record structure with data members \lstinline{Fields}.
A \lstinline{Field<Name, Type>} needs two type arguments.
The \lstinline{Name} type is used as a compile time identifier to refer to the field.
The \lstinline{Type} type is either an elemental type not further decomposed by LLAMA or another \lstinline{Record}.
If \lstinline{Type} is an array type with static extent, LLAMA replaces it by a \lstinline{Record} with as many \lstinline{Field}s as the array's extent.
The array dimensions are an N-dimensional array of runtime integral values, with N being a compile time value.
Listing~\ref{lst:data_structure_definition} shows an example of a data structure description in LLAMA.
For comparison, listing~\ref{lst:data_structure_definition_cpp} shows the equivalent data structure description in C++.

\noindent\begin{minipage}{.60\textwidth}
\begin{lstlisting}[caption=A data structure description in\\LLAMA., label=lst:data_structure_definition]
struct X{}; struct Y{}; struct Id{};
struct Pos{}; struct Mass{}; struct Flags{};

using Vec = llama::Record<
    llama::Field<X, float>,
    llama::Field<Y, float>
>;
using Particle = llama::Record<
    llama::Field<Id, uint16>,
    llama::Field<Pos, Vec>,
    llama::Field<Mass, double>,
    llama::Field<Flags, bool[3]>
>;
using ArrayDims = llama::ArrayDims<3>;
...
ArrayDims arrayDimsSize{128, 256, 32};
\end{lstlisting}
\end{minipage}\hfill
\begin{minipage}{.39\textwidth}
\begin{lstlisting}[caption=A data structure description in C++., label=lst:data_structure_definition_cpp]



struct VecCpp {
    float x;
    float y;
};
struct ParticleCpp {
    uint16 id;
    VecCpp pos;
    double mass;
    bool flags[3];
};



\end{lstlisting}
\end{minipage}

For a LLAMA data structure definition empty \lstinline{struct}s, called tag types, are created to serve as compile time names.
Then, a structure is defined via an alias declaration using the \lstinline{Record} construct, containing 2 \lstinline{Field}s of type \lstinline{float}, using the tag types as field names.
A second structure with 4 fields is then defined in a similar way, referring to the first structure for the second field, thus nesting the first structure.
Notice that for the last field a static array is specified.

\subsection{View creation}

The \lstinline{View} takes coordinates in the array and record dimensions and returns a reference to a record in memory which can be read from or written to.

As shown in listing~\ref{lst:view_creation}, a view is allocated using the helper function  \lstinline{allocView}, which takes a mapping and an optional blob allocator.

\begin{lstlisting}[caption=Creating a LLAMA view., label=lst:view_creation,escapechar=\$]
using Mapping = ...; // see section $\ref{sec:mappings}$
Mapping mapping(arrayDimsSize);
auto view = allocView(mapping, blobAlloc); // blobAlloc is optional
\end{lstlisting}

\subsection{Data access and VirtualRecord}

We call an access to a data structure which returns a memory location and the type stored at this location, a terminal access.
A non-terminal access on a data structure does not yet result in an element of the data structure but a subpart, an intermediate object.
The type of this intermediate object and how and if it materializes is a crucial design aspect of a data structure.

Traditional data structure access on C++ containers is resolved eagerly and the destination memory location is computed step-by-step.
E.g. the access \lstinline{std::vector<VecCpp> v; auto& x = v[i].x;} involves resolving to a \lstinline{VecCpp&} inside the \lstinline{std::vector} and then resolving \lstinline{x} on a \lstinline{VecCpp&} in two separate steps.
The \lstinline{VecCpp&} is an intermediary construct and already requires partial computation of the destination memory location.

In LLAMA, access to elements is built up gradually and only resolved once the access is terminal.
LLAMA therefore has a holistic view of an access request into a data structure and thus richer possibilities for optimization.
Listing~\ref{lst:access_vector_vs_llama} compares the access on a \lstinline{std::vector} and on a LLAMA \lstinline{View}:

\begin{lstlisting}[caption={
		Accessing a \lstinline|std::vector| vs. LLAMA view.
		Every non-terminal access results in a \lstinline{VirtualRecord} aggregating the already specified access index information.
		Every terminal access will invoke the Views mapping function to resolve to a blob and offset from where to serve the storage for a value.
	},label=lst:access_vector_vs_llama]
std::vector<ParticleCpp> vec = ...;
size_t i = 1 * pitch1 + 2 * pitch2 + 3; // index linearization
ParticleCpp& particle = vec[i];         // partial address computation
VecCpp& pos = particle.pos;             // partial address computation
float& y = pos.y;                       // final address computation

auto view = llama::allocView(...);
llama::VirtualRecord<...> particle = view(1, 2, 3); // non-terminal
llama::VirtualRecord<...> pos = particle(Pos{});    // non-terminal
float& y = pos(Y{});                // terminal, full address computation
\end{lstlisting}

A \lstinline{VirtualRecord} obtained from a \lstinline{View} is a value type that represents a reference.
\lstinline{One<RecordDim>} is a mechanism to create a view independent \lstinline{VirtualRecord}, which is a value type with deep-copy semantic.
Listing~\ref{lst:one} demonstrates the usage of \lstinline{One<RecordDim>}.

\begin{lstlisting}[caption={Using LLAMA's \lstinline{One} to create a local, view independent record value.},label=lst:one]
llama::One<Particle> p = view(3, 2, 1); // view independent deep-copy
p(Pos{})(Y{}) = 1.0;
auto p2 = p;                            // another independent deep-copy
view(1, 2, 3) = p2;                     // copy into view
\end{lstlisting}

\lstinline{VirtualRecord} overloads assignment, arithmetic and logical operators to allow to express computations on multiple fields.
For reference \lstinline{VirtualRecord}s, these operators directly write through into the corresponding view.
Binary operators return their results on the stack as a \lstinline{llama::One}.
These operators are designed to work even between two virtual records of different record dimensions, applying the operation on hierarchically matching tags.
Scalar operands are also supported, e.g. assigning to all record fields in one statement.
Since one operator executes on multiple independent fields, vertical vectorization\footnote{
	Vertical vectorization packs computations on multiple, different, scalar variables into a single SIMD operation.
	Conversely, horizontal vectorization packs multiple array instances of the same field into a SIMD operation.
} could be employed here, but is currently not implemented.

\subsection{C++ integration}

A \lstinline{VirtualRecord} models a reference to a structure with fields stored distributed in memory.
It behaves similarly to a \lstinline{std::tuple} holding references to the structure's fields.
LLAMA thus provides functionality, as demonstrated in listing~\ref{lst:integration}, to seamlessly convert between a structurally equivalent \lstinline{std::tuple}, \lstinline{VirtualRecord} and even a user defined data type if it implements the C++ tuple interface.
The latter is provided by specializing \lstinline{std::tuple_size<T>} and \lstinline{std::tuple_element<I, T>}, and providing a function \lstinline{get<I>(T)} for the user defined data type \lstinline{T}.
Structured bindings of \lstinline{VirtualRecord} are also supported.
In all cases, the mapping function is just invoked for terminal accesses.

\begin{lstlisting}[caption={Integration of \lstinline|VirtualRecord| with tuple like objects.},label=lst:integration]
// std::tuple_...<ParticleCpp> + get<I>(ParticleCpp)

auto p = view(1, 2, 3);
auto [id, pos, mass, flags] = p; // pos and flags are VirtualRecords
                                 // id is uint16& and mass is float&
auto [x, y] = pos;               // x, y are float&

ParticleCpp p1 = p.load(); // constructs ParticleCpp from multiple float&
p1.mass = 4.0;
p.store(p1);               // tuple-element-wise assignment from p1 to p
\end{lstlisting}

LLAMA offers several constructs to iterate over the record and array dimensions as demonstrated in listing~\ref{lst:iteration_constructs}.
\lstinline{ArrayDimsIndexRange} is a C++ range to iterate over all index tuples within the given array dimensions.
Compile-time iterating over the record dimension is implemented with \lstinline{forEachLeaf()}, passing different \lstinline{RecordCoord} types to the provided generic callable.

Iterators on views of any dimension are supported and allow LLAMA to integrate with the standard library.
Since virtual records interact with each other based on the record dimension tags, we can also use iterators from multiple views together, even if the views have different mappings or record dimensions.

\begin{lstlisting}[caption=LLAMA's iteration constructs and STL integration.,label=lst:iteration_constructs]
llama::ArrayDimsIndexRange range{llama::ArrayDims<2>{3, 3}};
std::for_each(range.begin(), range.end(), [](llama::ArrayDims<2> coord) {
    // coord is {0, 0}, {0, 1}, {0, 2}, {1, 0}, ..., {2, 2}
});
llama::forEachLeaf<Particle>([&](auto coord) {
    // coord is RecordCoord<0, 0>{}, ..., RecordCoord<3, 3>{}
    auto fieldValue = view(1, 2, 3)(coord);
});

for (auto p : view) p(Mass{}) = 1.0;

auto view2 = llama::allocView(...); // different layout, Vec as record dimension
std::transform(view.begin(), view.end(), view2.end(),
    [](auto p) { return p(Pos{}) * 2; });
auto [xsum, ysum] =
    std::reduce(view2.begin(), view2.end(), llama::One<Vec>{});
\end{lstlisting}

\subsection{Mappings}
\label{sec:mappings}

\begin{figure}
	\centerline{\includegraphics[width=\textwidth]{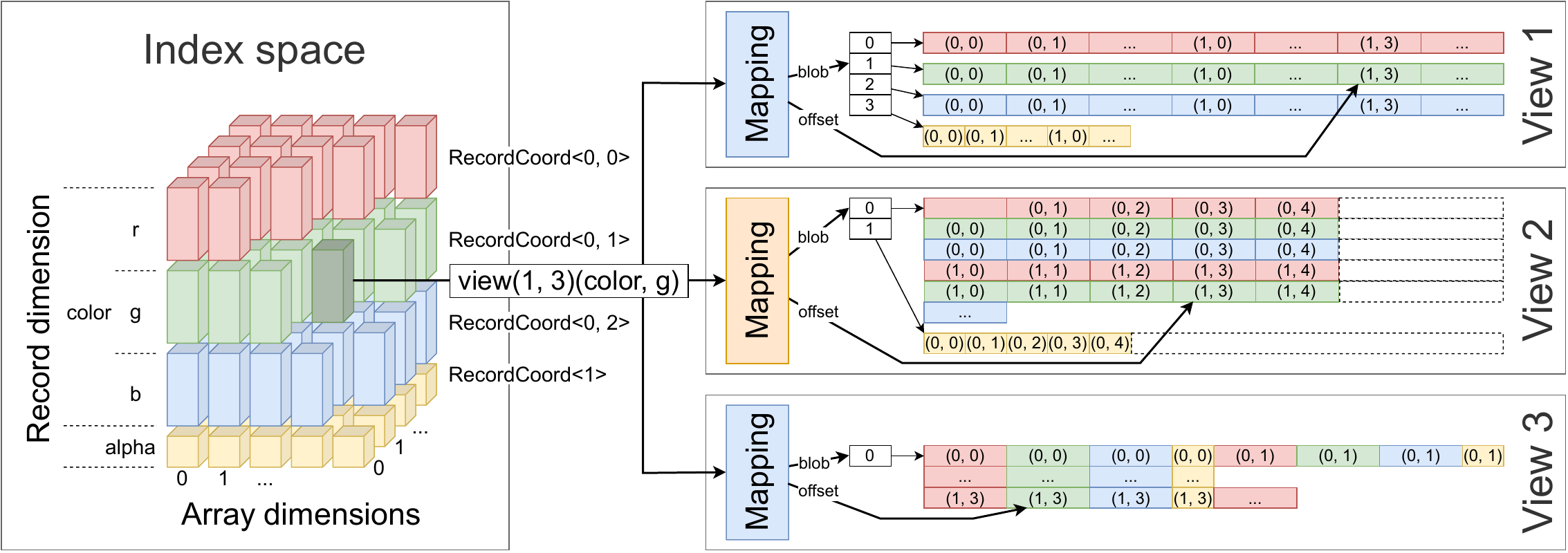}}
	\caption{
		Concept of LLAMA mappings.
		A mapping defines how a record and array dimension index tuple is translated into a blob number and offset.
		\label{fig:mapping}
	}
\end{figure}

The core task of LLAMA is to map an index tuple from the data/index space spanned by array and record dimensions to some address in the allocated memory space, as sketched in figure~\ref{fig:mapping}.
It is vital that the compiler can see through LLAMA's abstractions and into a mapping's implementation in order to optimize the memory accesses (vectorization, reordering, aligned loads, etc.).
Thus, mappings are compile time parameters to LLAMA's views.

Mappings are configured on the array and record dimensions.
They expose a compile time blob count, informing the view how many blobs it needs to hold.
Furthermore, for each blob the size in bytes can be queried at runtime.
The core functionality is the member function \lstinline{blobNrAndOffset<RecordCoords...>(ArrayDims) -> [blob, offset]}, which translates an access request to the element described by the \lstinline{RecordCoords} at an array dimensions coordinate into a destination blob number and offset inside that blob.
The view then takes this information to retrieve the actual byte address of the stored value and reinterpret it with the correct type.
Mappings can also be used for instrumentation and memory profiling.

LLAMA provides several ready-to-use mappings, but users are also free to supply their own mappings as long as they fulfill the \lstinline{Mapping} concept.
To demonstrate that implementing a mapping is often achievable with modest effort, we also list the LOCs of their implementations\footnote{As counted by the \lstinline[language=bash]{cloc} command.}, which use a few common utility functions not counted.
The following mappings are provided by LLAMA:

\begin{description}
	\item[AoS]
	48 LOCs.
	Places the record dimension's fields after each other into memory and repeats this layout as many times as the array dimensions have records.
	The fields can be padded to adhere to their type's alignment or tightly packed.

	\item[SoA]
	77 LOCs.
	For each field of the record dimension, repeats as many instances of one field contiguously as records in the array dimensions.
	Either one blob per field (called SoA MB, for multi-blob) or one blob for the entire layout can be used.

	\item[AoSoA]
	61 LOCs.
	Same as AoS, but repeats a field $L$ times before continuing with the next field in the record domain.
	$L$ is a compile time argument of this mapping.

	\item[One]
	34 LOCs.
	Collapse the entire array dimensions into 1 element, thus storing a single instance of the record domain.

	\item[Split]
	139 LOCs.
	Selects a part of the record domain by a record coordinate and splits the record domain, mapping the selected part using a different mapping than the remaining part.

	\item[Trace]
	72 LOCs.
	Counts the number of accesses to each field at runtime and then forwards the access to an inner mapping.
	This information can be printed to help a user understand the access behavior of their program.

	\item[Heatmap]
	60 LOCs.
	Similar to the Trace mapping, but counts the number of accesses to individual bytes.
	The resulting information can be plotted as heatmap of the memory.
\end{description}

These mappings are built using a few LLAMA building blocks which can also aid users in implementing their mappings, e.g. functions to linearize the array dimensions, computing sizes and offsets in the record dimension or type list algorithms to permute the record dimension to minimize padding introduced by alignment.

\begin{figure}
	\subfloat[Packed AoS mapping.]{
		\includegraphics[width=0.47\textwidth]{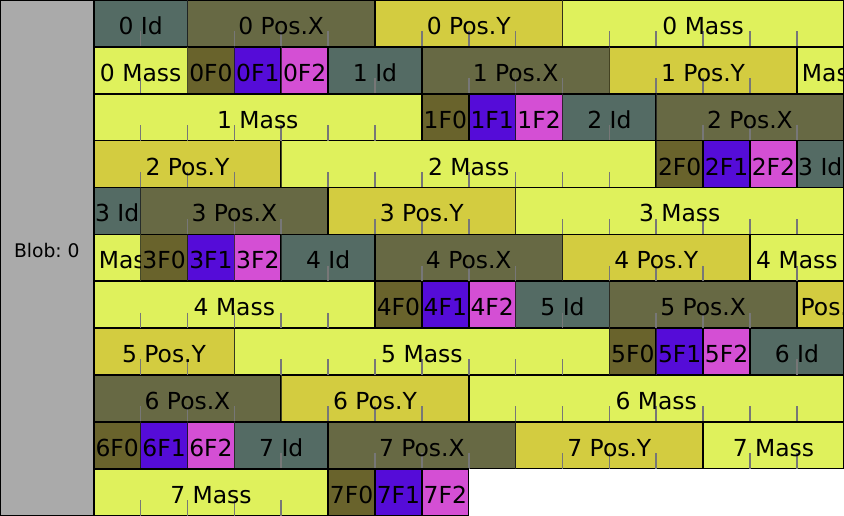}
	}
	\hfil
	\subfloat[AoSoA4 mapping.]{
		\includegraphics[width=0.47\textwidth]{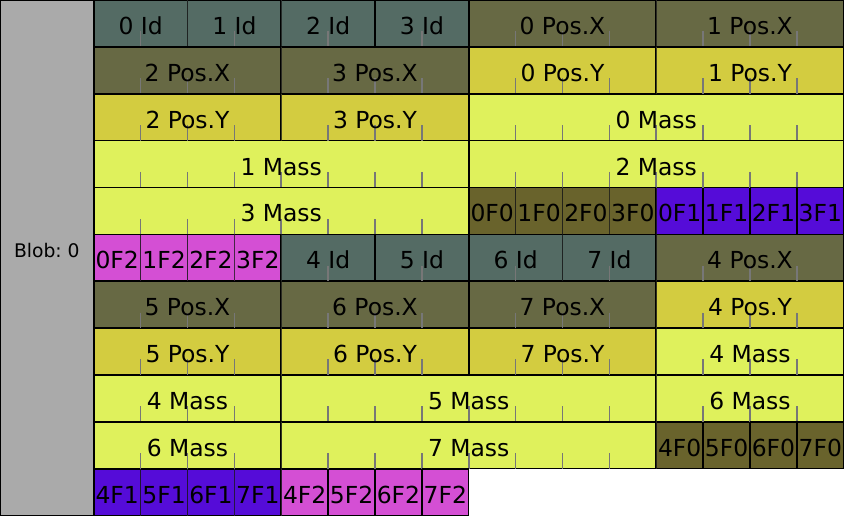}
	}
	\newline
	\subfloat[Split mapping, where the field at record coordinate 1 (Pos) is split off into a multiblob SoA, followed by splitting the new record coordinate 1 (Mass) into the mapping One and layouting the remaining record (Id and Flags) as aligned AoS.]{
		\includegraphics[width=0.47\textwidth]{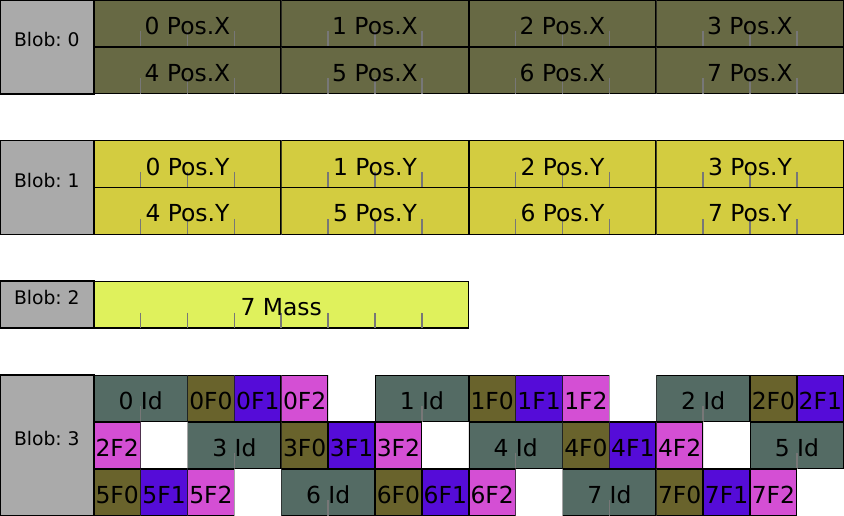}
	}
	\hfil
	\subfloat[Heatmap showing the number of accesses to the bytes of an AoS layout during a small run of the n-body simulation discussed in section~\ref{sec:nbody}.]{
		\includegraphics[width=0.47\textwidth]{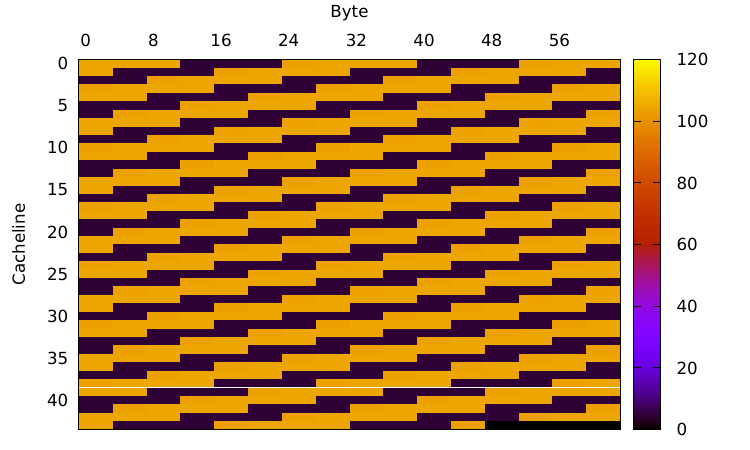}
	}
	\caption{
		Three examples of different mappings of the same data structure and a heatmap as they are created by LLAMA.
		The text in the blocks containing the \lstinline{Flags} field has been manually shortened.
		\label{fig:mapping_examples}
	}
\end{figure}

Listing~\ref{lst:mapping_examples} shows how to specify a 3D AoS and other more complicated mappings in LLAMA and how to dump an SVG image of them.
A flexible HTML visualization can also be dumped but is not shown in this article.
The corresponding SVG dumps and the heatmap plot of the memory layouts are shown in figure~\ref{fig:mapping_examples}.

\begin{lstlisting}[caption={Creating a LLAMA AoS mapping, an AoSoA mapping with 4 lanes, a combination of various nested mappings by splitting the record domain, and adding instrumentation to a mapping producing a heatmap.},label=lst:mapping_examples]
using MappingA = llama::mapping::AoS<ArrayDims, Particle>;
using MappingB = llama::mapping::AoSoA<ArrayDims, Particle, 4>;
using MappingC = llama::mapping::Split<ArrayDims, Particle,
        llama::RecordCoord<1>,
        llama::mapping::PreconfiguredSoA<true>::type,
        llama::mapping::PreconfiguredSplit<llama::RecordCoord<1>,
            llama::mapping::One, llama::mapping::AlignedAoS, true>::type,
    true>
auto mapA = MappingA{arrayDimsSize};          // analogous for MappingB/C
auto mapD = llama::mapping::Heatmap{mapA};
std::ofstream{"a.svg"} << llama::toSvg(mapA); // analogous for mapB/C
std::ofstream{"d.sh"} << mapD.toGnuplotScript();
\end{lstlisting}

\subsection{Blobs}

When constructed, a \lstinline{View} needs an array of blobs to store its data.
A blob is an object representing a contiguous region of memory where each byte is accessible using the subscript operator.
This allows views to operate on a broad range of memory sources such as owning containers, smart pointers, non-owning constructs like \lstinline{std::span<std::byte>}, raw pointers, memory mapped files, pointers to CUDA global/shared/constant memory, etc.
The separation of providing allocated memory as blobs and interpreting the memory inside view and mapping allows LLAMA to function orthogonally to memory allocation.
Blobs for a view are either allocated via \lstinline{allocView()}'s second parameter, a callable named blob allocator, or by passing an array of blobs directly to a view's constructor.

\subsection{Copying between views}

Especially when working with hardware accelerators such as GPUs, or offloading to many-core processors, explicit copy operations call for large, contiguous memory chunks to reach good throughput.

Copying the contents of a view from one memory region to another if mapping and size are identical is trivial.
However, if the mapping differs, a direct copy of the underlying memory is wrong.
In many cases only field-wise copy operations are possible.

There is a small class of remaining cases where the mapping is the same, but the size or shape of the view is different, or the record dimension differ slightly, or the mappings are very related to each other.
E.g. when both mappings use SoA, but one time with, one time without padding, or a specific field is missing on one side.
Or two AoSoA mappings with a different inner array length.
In those cases an optimized copy procedure is possible, copying larger chunks than mere fields.

For the moment, LLAMA implements a generic, field-wise copy with specializations for combinations of SoA and AoSoA mappings, reflect the properties of these.
This is sub-optimal, because for every new mapping new specializations are needed.

One thus needs new approaches on how to improve copying because LLAMA can provide the necessary infrastructure:
\begin{itemize}
	\item
	A run time analysis of the two views to find contiguous memory chunks.
	The overhead is probably big, especially if no contiguous memory chunks are identified.

	\item
	A black box compile time analysis of the mapping function.
	All current LLAMA mappings are \lstinline{constexpr} and can thus be run at compile time.
	This would allow to observe a mappings behavior from exhaustive sampling of the mapping function at compile time.

	\item
	A white box compile time analysis of the mapping function.
	This requires the mapping to be formulated transparently in a way which is fully consumable via meta-programming, probably at the cost of read- and maintainability.
	Potentially upcoming C++ features in the area of statement reflection\footnote{
		Statement reflection in C++ is the proposed ability to reflect arbitrary C++ statements at compile time, making their abstract syntax tree (AST) available for meta-programming.
		This ability is discussed as part of automatic differentiation~\cite{autodiff_cpp} and lazy expression evaluation~\cite{reflection_based_lazy_eval}.
	} could improve these a lot.
\end{itemize}

Copies between different address spaces, where elementary copy operations require calls to external APIs, pose an additional challenge and profit especially from large chunk sizes.
A good approach could use smaller intermediate views to shuffle a chunk from one mapping to the other and then perform a copy of that chunk into the other address space, potentially overlapping shuffles and copies in an asynchronous workflow.

\section{Evaluation}
\label{sec_evaluation}

\newcommand{\speccpu}{SPEC CPU\textsuperscript{\textregistered}\xspace}

The evaluation benchmarks were selected to show various aspects of LLAMA and effects of the memory layouts.
We present an n-body simulation containing a compute intensive and memory intensive computation with quadratic and linear runtime complexity.
This benchmark reflects classes of codes that benefit from maximizing data throughput to minimize time to solution and thus maximize compute performance.
Furthermore, we include a layout changing copy benchmark, showing pure shuffling of data with no computation.
This example from HEP focuses on throughput and looks at complex data structures for the storage of elementary particle data usually found in large-scale data analysis of detector data.
Here, the goal is optimizing copies between memory layouts.
To demonstrate LLAMA's use in real-world applications, we show integrations into the \speccpu lbm benchmark and into the particle-in-cell simulation PIConGPU.

The n-body CPU, layout changing copy and \speccpu lbm benchmark were run on systems with an 8 core Intel i7-7820X with 32 GB RAM and on an 64 core AMD EPYC 7702 with 512 GB RAM.
They were compiled with g++ 10.2 on the Intel system and g++ 11.1 on the AMD system.
The compiler flags were \lstinline[language=bash]{-O3 -march=native -ffast-math -fopenmp}, except for the n-body on the Intel system, where we used \lstinline[language=bash]{-mavx2 -mfma} instead of \lstinline[language=bash]{-march=native} to use the same compute instruction set as the AMD system, and because Vc does not support AVX512.
Multi-threading was done via GNU OpenMP with as many pinned threads as cores (no hyper-threading).
The n-body GPU benchmark was run on systems with an Nvidia TITAN Xp 12GB, an Nvidia Quadro RTX 8000 48GB and an Nvidia Tesla V100-SXM2 32GB.
It was compiled with g++ 9.3/nvcc 11.3, g++ 8.3.1/nvcc 11.0 and g++ 10.2/nvcc 11.2, respectively, with no specific flags for g++ and the flags \lstinline[language=bash]{-O3 --generate-code=arch=compute_35,code=[compute_35,sm_35] --expt-extended-lambda --expt-relaxed-constexpr --use_fast_math} for nvcc.
We tried setting higher CUDA architecture versions as well for nvcc (i.e. \lstinline{sm_70} for the V100), but that did not affect the results.
The PIConGPU simulation was run on the same Nvidia Tesla V100 system as the n-body GPU benchmark.
All reported numbers are the average of 5 consecutive runs.

\subsection{All pairs n-body simulation}
\label{sec:nbody}

N-body simulations occur in a multitude of scientific domains, including protein folding, astrophysics or fluid dynamics.
An n-body simulation computes the interaction between a set of particles subject to a force.
The simulation runs in time steps consisting of two phases.
Firstly, each particle's 3D velocity is updated based on the influence of all other particles' 3D positions and masses.
This phase is called \lstinline{update} and is compute intensive.
Secondly, each particle's position is updated based on its velocity.
This phase is called \lstinline{move} and is memory intensive.

We investigate a single threaded implementation here because we want to focus on the effects of local data representation.
Furthermore, the n-body update usually consumes more time on compute than what is required for memory transfers between nodes.
Runs with multiple threads on a shared memory machine have been conducted and show the expected scaling, which is linear for the compute-bound \lstinline{update}, and starts linear for \lstinline{move}, flattening with higher thread counts because the memory subsystem saturates.

Listing~\ref{lst:nbody_llama_short} shows part of a LLAMA implementation of an n-body simulation:

\begin{lstlisting}[caption=LLAMA implementation of an n-body simulation. The full version is available in LLAMA's GitHub repository.,label=lst:nbody_llama_short]
constexpr auto PROBLEM_SIZE = 16 * 1024;
using FP = float;
constexpr FP TIMESTEP = 0.0001f, EPS2 = 0.01f;

void pPInteraction(auto& pi, auto pj) {
    auto dist = pi(Pos{}) - pj(Pos{}); dist *= dist;
    const FP distSqr = EPS2 + dist(X{}) + dist(Y{}) + dist(Z{});
    const FP distSixth = distSqr * distSqr * distSqr;
    const FP invDistCube = 1.0f / std::sqrt(distSixth);
    const FP sts = pj(Mass{}) * invDistCube * TIMESTEP;
    pi(Vel{}) += dist * sts;
}

void update(auto& particles) {
    LLAMA_INDEPENDENT_DATA // == #pragma ivdev
    for (std::size_t i = 0; i < PROBLEM_SIZE; i++) {
        llama::One<Particle> pi = particles(i);
        for (std::size_t j = 0; j < PROBLEM_SIZE; ++j)
            pPInteraction(pi, particles(j));
        particles(i) = pi;
    }
}

void move(auto& particles) {
    LLAMA_INDEPENDENT_DATA
    for (std::size_t i = 0; i < PROBLEM_SIZE; i++)
        particles(i)(Pos{}) += particles(i)(Vel{}) * TIMESTEP;
}
\end{lstlisting}

\begin{figure}
	\subfloat[Update with 64ki particles]{
		\includegraphics[width=0.49\textwidth]{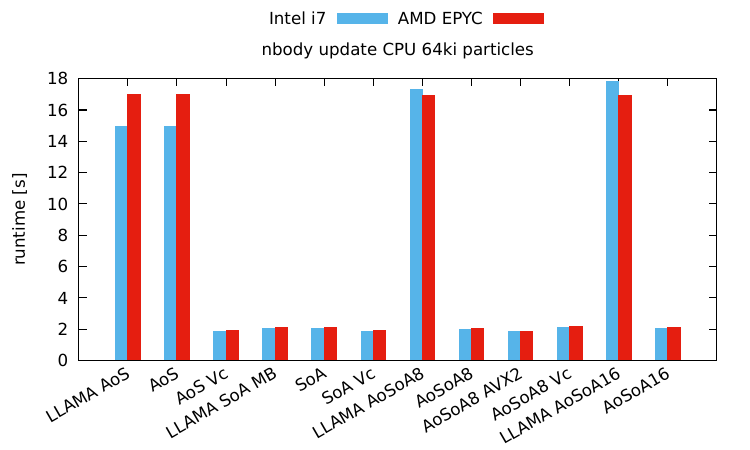}
	}
	\hfil
	\subfloat[Move with 256Mi particles]{
		\includegraphics[width=0.49\textwidth]{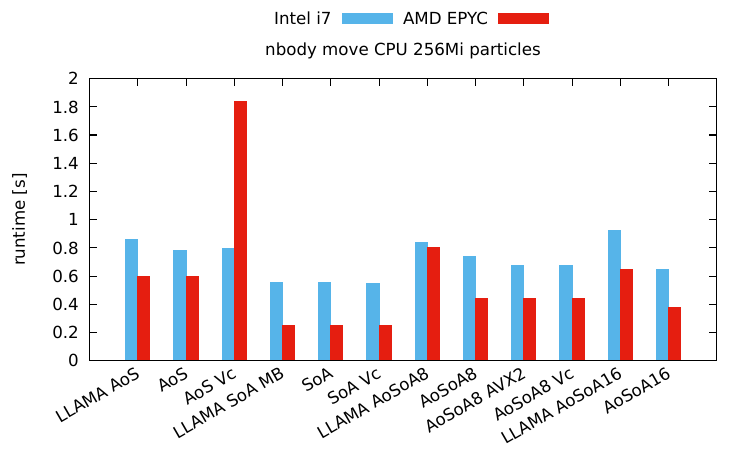}
	}
	\caption{
		Runtime comparison of n-body CPU implementations using various memory layouts and technologies.
		We chose a different problem size for update and move, because the former scales quadratically with the problem size and the latter only linearly.
		The runtime difference between AoS and multi-blob SoA (SoA MB) layouts is clearly visible, and both LLAMA layouts perform as their manual implementations.
		LLAMA's AoSoA suffers from an unfavorable loop structure not being vectorized, compared to the manually written AoSoA versions.
		\label{fig:nbody_cpu_titanx}
	}
\end{figure}

Figure~\ref{fig:nbody_cpu_titanx} shows the runtime of the \lstinline{update} and \lstinline{move} functions with various LLAMA and non-LLAMA implementations on the Intel and AMD CPU.
We can see that the versions with LLAMA AoS or SoA mapping perform almost the same as their manually written versions.
This confirms that LLAMA provides a zero-overhead abstraction with these mappings.

For \lstinline{update}, the runtime difference between AoS and SoA layouts is clearly visible.
However, a large degree of the runtime difference comes from the compiler being able to auto vectorize the SoA version.
In fact, the disassembly of the LLAMA and manual AoS/SoA implementations is entirely the same (cf. SoA disassembly in listings~\ref{lst:manual_soa_disassembly} and~\ref{lst:llama_soa_disassembly} in the appendix), showing just how well the compiler is able to see through LLAMA's abstractions.
Therefore, we also include a manually vectorized AoS version (AoS Vc) using gather and scatter instructions, which is equally fast than the SoA versions.
This shows that the memory layout has little impact on the \lstinline{update} function, since most time is spent in computation.
Still, creating the manually vectorized AoS version is additional effort.
Maybe LLAMA can help here in the future by providing explicit integration with SIMD libraries.

LLAMA AoSoA layouts perform comparably poor.
This stems entirely from the design of view indexing and the AoSoA mapping implementation.
As seen in listing~\ref{lst:nbody_llama_short}, the \lstinline{update} function uses a single loop with one induction variable for a single pass over the view's particles.
Internally, an AoSoA$L$ mapping needs to split the loop index $i$ into the inner and outer array indices like $i \mapsto (\frac{i}{L}, i \bmod L)$.
As $L$ is known at compile time and usually a power of two, this split results in a bit shift and mask operation, which prevents the compiler from vectorizing and also adds additional overhead.
This problem does not occur in the manually written AoSoA versions, as these use two nested loops for a single pass over the particles, with the inner loop having a trip count of $L$, allowing the compiler to fully unroll and vectorize it.
To account for this, LLAMA would need a dedicated iteration mechanism, allowing nested loops, which is aware of the mapping's needs.
Further investigation is pending here.
For reference, we also included a manually written AoSoA version with explicit AVX2 intrinsics and a manually written SoA and AoSoA using the Vc SIMD library (using AVX2 as well).

For \lstinline{move}, we can nicely see the difference between AoS and SoA.
From a particle's 7 floats, 6 are read and 3 are written.
Since the memory subsystem deals in cache lines, the AoS generally wastes $\frac{1}{7}$ on loads and $\frac{4}{7}$ on stores, which amounts to $1-\frac{1+4}{7+7} = 64.3\%$ use of bandwidth to transport useful data.
The SoA layout mitigates this issue by separating the storage regions of the particle's fields and the full bandwidth is meaningfully used, resulting in the consumption of only 64.6\% of the AoS's runtime.
Thus the choice of memory layout has a large impact.
A manual vectorization of the AoS does not yield any benefit, as the routine's compute consists only of 3 FMAs and loads/stores inside a loop, making it memory bound.
For the AMD CPU, the AoS gathering and scattering even introduces a large overhead, making it multiple times slower as a scalar AoS.
LLAMA's AoSoA here suffers from the same bit shift and mask overhead as the \lstinline{update} routine.
For reference, we also included a manually written and vectorized versions of SoA and AoSoA version using the Vc SIMD library.

\begin{figure}
	\subfloat[Update with 1024ki particles]{
		\includegraphics[width=0.49\textwidth]{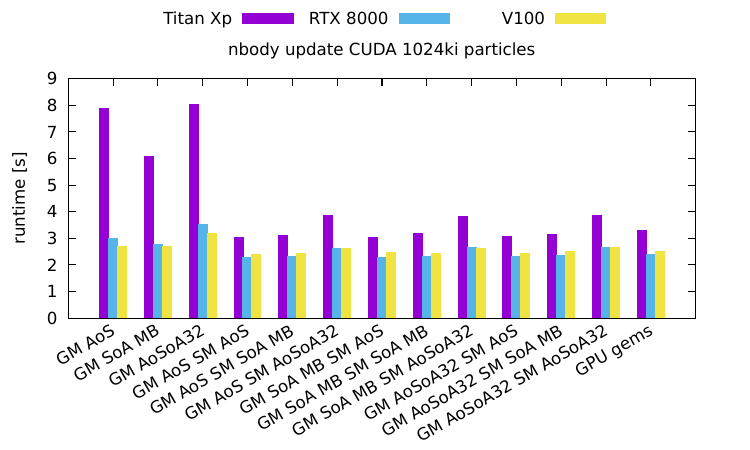}
	}
	\hfil
	\subfloat[Move with 256Mi particles]{
		\includegraphics[width=0.49\textwidth]{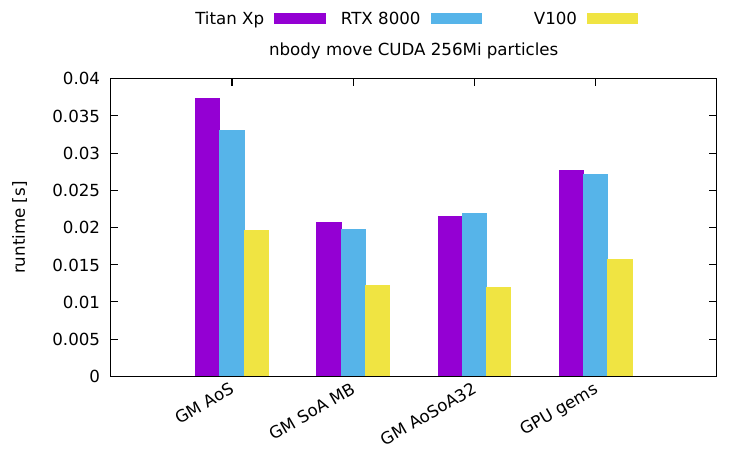}
	}
	\caption{
		Runtime comparison of n-body CUDA implementations using various memory layouts.
		GM denotes the memory layout used for global memory.
		Updates with SM use shared memory for tiled caching with the specified layout.
		For kernels without shared memory, the choice of memory layout has a stronger impact, with the multi-blob SoA (SoA MB) layout performing best.
		Kernels with shared memory on less powerful GPUs perform generally better, but the global memory layout is less important.
		The AoSoA for shared memory adds some overhead.
		Moving the particles is similarly fast with SoA and AoSoA.
		For reference, the n-body implementation from GPU Gems 3~\cite{gpu_gems_3} has been included.
		\label{fig:nbody_gpu}
	}
\end{figure}

Figure~\ref{fig:nbody_gpu} shows the runtime of the \lstinline{update} and \lstinline{move} functions with various LLAMA implementations on the GPU using CUDA.
The kernels are launched with 256 threads per block.
Kernels with shared memory cache 512 particles (7kiB) per block in CUDA shared memory, allowing sufficient occupancy of the warps.
Tests with other numbers of cached particles revealed equal or worse performance.

For \lstinline{update}, the first three runs avoid shared memory for caching so the effect of the memory layout is fully visible.
The SoA layout performs visibly best here for the Titan Xp.
For the stronger Quadro and V100, the global memory layout seems to have less impact.
Even though the AoSoA with 32 lanes should allow coalesced memory access, it is slightly slower than the AoS layout.
For the variants using shared memory, the choice of global memory layout is largely irrelevant.
Also, whether AoS or SoA is used for shared memory has almost no impact.
AoSoA for the shared memory has a noticeable impact, but not a grave one.
We assume this is again due to the increased cost of evaluating the mapping function, as we can see in the disassembly.

For \lstinline{move}, similarly to the CPU variants, we can also see the difference between AoS and SoA.
The SoA variant requires 55/60/62\% of the AoS runtime and therefore shows both, better use of the loaded data (the AoS only uses 64.3\% of the loaded data) and better utilization of the memory subsystem.
Again, the choice of memory layout has a large impact here.
The AoSoA layout performs slightly slower than the SoA.

We still lack deeper insight into the architectural implications of the various memory layouts.
In the disassembly we could observe that SoA mappings require one register per record field, whereas AoS just requires a single register for the entire view.
This might be relevant for large record dimensions.
From profiling we further observe that different memory layouts cause vastly different levels of utilization of the compute and memory subsystems, although the overall runtime ends up quite similar.
A possible explanation could be that the n-body problem as presented here is not demanding enough.
This is subject to further investigation.

\subsection{Layout changing copying}

In the future data throughput will depend largely on the performance of copy operations.
This includes streaming data through the memory hierarchy and shuffling data for better parallel access.
In the following benchmark we will investigate the runtime of copying two large data sets from one memory layout to another.
Whereas the n-body benchmark in section~\ref{sec:nbody} performed additional computations, the benchmark presented now purely shuffles data.

\begin{figure}
	\centerline{\includegraphics[width=\textwidth]{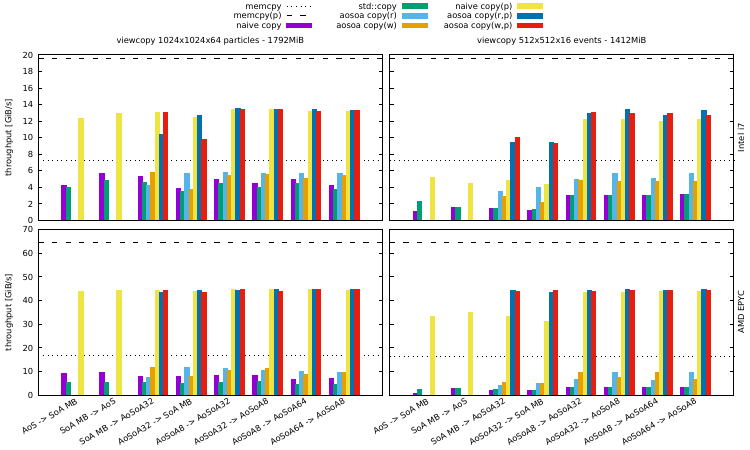}}
	\caption{
		Throughput comparison of several copy implementations between different mappings.
		The particle record dimension consists of 7 \lstinline{float}s.
		The event record dimension consists of the first 100 \lstinline{int32}s, \lstinline{int64}s, \lstinline{float}s, \lstinline{byte}s and \lstinline{bool}s as they occur in an internal event dataset from the CMS detector at CERN (cf. the example on LLAMA's GitHub repository for the full definition).
		The (p) versions are parallel.
		The (r) versions read contiguously, whereas the (w) versions write contiguously.
		For the field-wise naive and \lstinline{std::copy}, the throughput depends a lot on the field types of the record dimension.
		The \lstinline{aosoa_copy} outperforms both of them in most cases, single- and multithreaded.
		\label{fig:viewcopy_cpu_titanx}
	}
\end{figure}

Figure~\ref{fig:viewcopy_cpu_titanx} shows the throughput achieved when copying 7 field particles and 100 field records of HEP event data\footnote{
	Events in a HEP detector are usually considered the result of an interaction.
	One such interaction can result in a very large number of events.
	An event typically contains a multitude of detector measurements and reconstructed information from those.
} between two views of using different memory layouts using several strategies.
The \lstinline{naive copy} consists of nested loops over the array and record dimensions and copies field-wise.
The \lstinline{std::copy} version uses the equally named STL algorithm and iterators on the LLAMA view, also copying field-wise.
The \lstinline{aosoa_copy}s are specialized and layout-aware copy algorithms that copy $min(N,M)$ fields as chunks between an AoSoA$N$ and an AoSoA$M$ layout.
While copying chunks, the (r) versions of \lstinline{aosoa_copy} read contiguously, whereas the (w) versions write contiguously.
SoA can be seen as an AoSoA with an inner array length equal to the product of the array dimensions.
For reference, the throughput of \lstinline{memcpy} with one and multiple threads is also listed.

The throughput of the field-wise naive copy turns out better than anticipated for simple record dimensions, as seen when copying the particles, even outperforming one version of the specialized \lstinline{aosoa_copy} when copying particles between SoA MB and AoSoA32.
By contrast, when copying the heterogeneous event data, the element-wise copies perform significantly worse.
We assume this is due to the loop structure, which iterates the record fields in the inner loop and thus access memory with a large stride on SoA MB and a medium stride on AoSoA layouts.

The parallel version of the naive field-wise copy for the particles is on a par with the \lstinline{aosoa_copy} and close to it when copying events, unless one of the layouts is a SoA MB, where the naive copy performs notably bad.
We again attribute this to the loop structure.

The \lstinline{std::copy} uses the LLAMA view's iterator and executes the same elemental, field-wise copies as the naive copy.
It is slightly slower in most cases because the iterators need to map the 1D iteration inside \lstinline{std::copy} to the 3 array dimensions (which are later linearized by the mapping again), introducing some overhead.
Only when copying events from AoS to SoA MB, \lstinline{std::copy} is significantly faster than the naive copy.
This circumstance is still subject to further investigation, but it seems to be related to a different unrolling done by the compiler.

The \lstinline{aosoa_copy} generally achieves higher throughput for AoSoA and SoA MB mappings, but is only applicable to such mappings.
Except for the SoA from/to AoSoA32 particles copy, the single threaded version consistently outperforms the field-wise copy, demonstrating its advantage of knowing the underlying mapping and copying in larger chunks.
In all our measurements, the parallel version has either the highest throughput or is among the best performing copy implementations.
There is a significant difference between the contiguously reading (r) and writing (w) versions of \lstinline{aosoa_copy} when copying particles between SoA MB and AoSoA32 on the Intel CPU, again subject to further investigation.

When inspecting the disassembly the compiler is generating scalar move instructions for the naive and \lstinline{std::copy}.
For the \lstinline{aosoa_copy} on the other hand, which can copy in larger blocks, the compiler generates packed vector move instructions.

\subsection{\speccpu 2017 lbm benchmark}

The Standard Performance Evaluation Corporation (SPEC)~\cite{spec_cpu_2017} is an organization providing a series of standardized benchmarks.
Among these, the \speccpu 2017 benchmark package measures speed and throughput of modern computing systems on integer and floating-point focused computations, thus providing ideal candidates for testing LLAMA's efficiency on real-world code.
However, with the integration of LLAMA into one of the benchmarks, we are no longer allowed to present a valid SPEC metric as we disobey several SPEC imposed run rules.
Specifically, we compile with C++17 enabled instead of C++03, not run our benchmark via the \lstinline[language=bash]{runcpu} benchmark driver and not as part of one of the \speccpu 2017 suites.
We report the runtime relative to the original benchmark, which again, is not a valid SPEC metric.

We chose the 619.lbm\_s benchmark~\cite{spec_cpu_2017_lbm} for our evaluation of LLAMA, because it has a single central data structure and a reasonable size for integration\footnote{
	\speccpu 2017 mostly includes complex benchmarks with many data structures, like compiling a large source file with the GNU C compiler or 3D rendering with Blender, with are unreasonably large to port to LLAMA within the scope of this article.
}.
This benchmark uses the Lattice Boltzmann method to simulate incompressible fluids in 3D.
The central data structure is a 3D array of 20 double precision values (of which one is used as bitset).
In addition to the flags already mentioned at the beginning of section~\ref{sec_evaluation} we further passed the definitions \lstinline[language=bash]{-DLARGE_WORKLOAD} and \lstinline[language=bash]{-DSPEC_OPENMP} to the compiler to enable a large grid size and OpenMP.
After a small refactoring, the LLAMA integration was straight forward, which we attribute to the existing separation of data access from the rest of the simulation code with preprocessor macros.
While this approach is limited to define only one access scheme for the entire application, it clearly shows that the benchmark authors were aware that data access needs to be decoupled from computation code so it can be independently adapted\footnote{
	Some discussion on this specific example is given by Thomas Pohl et al.\cite{lbm_paper}.
}.
We used LLAMA's array dimensions to describe the 3D array and the record dimension to describe the inner 20 double precision values.

\begin{figure}
	\centerline{\includegraphics[width=0.75\textwidth]{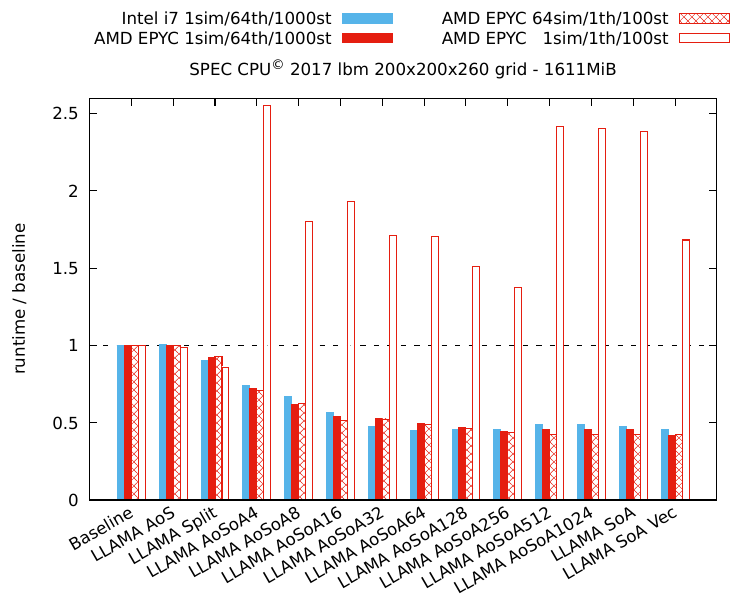}}
	\caption{
		Relative runtime comparison of the \speccpu 2017 619.lbm\_s benchmark using a 200x200x260 grid (1611MiB) with various modified versions using LLAMA.
		These are not SPEC metrics.
		For the 1sim benchmarks we run 1 simulation and report the pure compute time, not including the loading of the obstacle file, with 64 or 1 thread.
		The 64sim benchmark runs 64 simulations in parallel, each with their separate data, and includes loading the obstacle file.
		When the CPU cores are fully utilized (either 1sim/64th or 64sim/1th), the LLAMA AoS mapping is as fast as the original benchmark (Baseline), the Split mapping a bit faster.
		The AoSoA mapping becomes faster with increasing lane count and even beats the SoA mapping in some cases.
		Vectorization yielded only minor performance gains.
		When the CPU cores are not fully utilized (1sim/1th), the trend reverses and AoSoA and SoA are slower than the AoS and Split, with Split being the fastest one.
		\label{fig:lbm}
	}
\end{figure}

Figure~\ref{fig:lbm} shows our runtime measurements of the 619.lbm\_s benchmark and our modified versions using LLAMA.
At first, we will only discuss the multi-threaded benchmarks (64th).
The LLAMA AoS version is 0.4\% slower on the Intel and 0.2\% faster on the AMD CPU compared to the original benchmark (Baseline), thus showing negligible or no overhead of the LLAMA abstraction.
We temporarily wrapped the LLAMA Trace mapping around the LLAMA AoS mapping to inspect the number of memory accesses to each member of the record dimension.
Based on this result, we split the record dimension into 4 groups of AoS layouts with equal access count\footnote{Access counts are 83,055,344 on 1 member, 112,926,032 on 6 members and 137,949,048 on 12 members and 77,977,224 on the bitfield member.} using the LLAMA Split mapping, attempting a separation of the hot/cold data.
The performance gain of 9.5\% on the Intel and 7.6\% on the AMD CPU is marginal.

Switching to the LLAMA SoA (single-blob) was a matter of changing a single line of code and resulted in an impressive runtime reduction, taking 47.4\% of the baseline time on the Intel and 45.5\% on the AMD CPU.
The compiler was not able to automatically vectorize the code with the SoA layout due to several branches in the innermost loop.
Thus, the speedup comes solely from the different memory layout and not from using a different instruction set.
Within a multi-day effort, we could restructure the code to help the compiler produce a vectorized version using the SoA memory layout.
We pulled a loop over the SIMD lane count out of the innermost loop, rewrote the branches as conditional assignments to allow the compiler to generate conditional SIMD move instructions, added an early exit check whether all loops are processing an obstacle cell, and applied the \lstinline{LLAMA_INDEPENDENT_DATA} macro.
Using the vectorized version we could gain another 2\% on the Intel and 3.8\% on the AMD CPU over the non-vectorized SoA, which are 45.5\% and 41.6\% of the baseline runtime, respectively.
For the Intel CPU, supporting AVX512, we tested with 256bit and 512bit register sizes.
The runtime difference was 0.1\% compared to the baseline and we report the faster result using 512bit registers.

With the flexibility of LLAMA, we could also easily inspect the runtime of several versions of the AoSoA layout, which all produced scalar, non-vectorized code.
Starting from 4 we doubled the AoSoA lane count until 1024 and observed an initial steady increase in performance until flattening out at the runtime of the SoA layout.
The fastest AoSoA runs were achieved with 64 lanes on the Intel and with 256 lanes on the AMD CPU, scoring 44.8\% and 44.2\% of the baseline runtime.
Both runs outperformed the SoA layout by a difference of 2.6\% (Intel) and 1.3\% (AMD) compared to the baseline, and on the Intel CPU, the scalar AoSoA64 even outperformed the vectorized SoA with a difference of 0.7\%.

We also ran the benchmarks shown in figure~\ref{fig:lbm} with a single thread and OpenMP disabled (1th), thus underutilizing the CPU.
This might be a scenario for certain, inherently serial programs, or when reduced workloads are used for debugging and profiling.
In this case the LLAMA AoS and Split layouts were 1.6\% and 14.1\% faster than the baseline.
On the contrary, the AoSoA layouts were between 37.4\% (AoSoA256) and 154.9\% (AoSoA4) slower than the baseline.
Also, the SoA and vectorized SoA were 138.4\% and 68.2\% slower than the baseline.
When running as many processes of the single threaded benchmark (64sim/1th) as available cores in parallel (multi-process), and thus saturating the CPU again, the AoSoA and SoA outperform the AoS and Split again, showing strikingly similar runtime characteristics as the original multithreaded benchmark;
even though the benchmark now maintains 64 independent grids instead of one and includes the time to load the obstacle file.
We thus conclude that the optimal memory layout for a program is also greatly impacted by the use and saturation of the executing CPU by other processes.
This has to be kept in mind when e.g. debugging or profiling the behavior of a program on a reduced working set which does not fully utilize the system, as the results might give completely misleading conclusions.
It also seems that whether a program employs multithreading or multiprocessing to distribute work has only a negligible effect on finding the optimal memory layout.
However, this hypothesis would need further exploration using programs properly prepared for multiprocessing, like MPI based simulations.

\subsection{PIConGPU}

PIConGPU is a highly performant particle-in-cell simulation framework for CPUs, GPUs and many-core architectures~\cite{picongpu}.
It is built with a focus on making as much simulation specific configuration as possible available at compile time, including the precise definition of particle and field attributes.
The two main data structures used in PIConGPU are 2D or 3D staggered grids \cite{staggered_grid_yee} for the electric, magnetic, and current fields, and particle frame lists to store the attributes of particles which are currently inside a chunk of grid cells, called a supercell.

\begin{figure}
	\noindent\begin{minipage}{.49\textwidth}
			\centerline{\includegraphics[width=\textwidth]{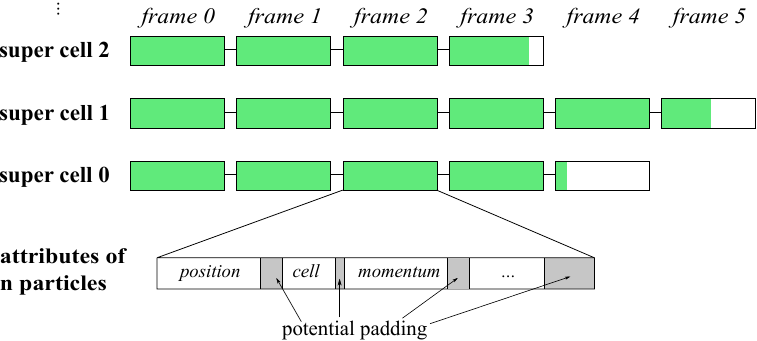}}
			\caption{
				The particle frame list data structure in PIConGPU.
				Supercells have linked lists of fixed-size particle frames containing particle attributes laid out as SoA.
				Padding might be inserted between the particle attribute arrays to fulfill the attribute type's alignment.
				Image adapted from Axel~Huebl~\cite{picongpu_axel_huebl} under CC-BY license.
				\label{fig:picongpu_frame_list}
			}
	\end{minipage}\hfill
	\begin{minipage}{.49\textwidth}
			\centerline{\includegraphics[width=\textwidth]{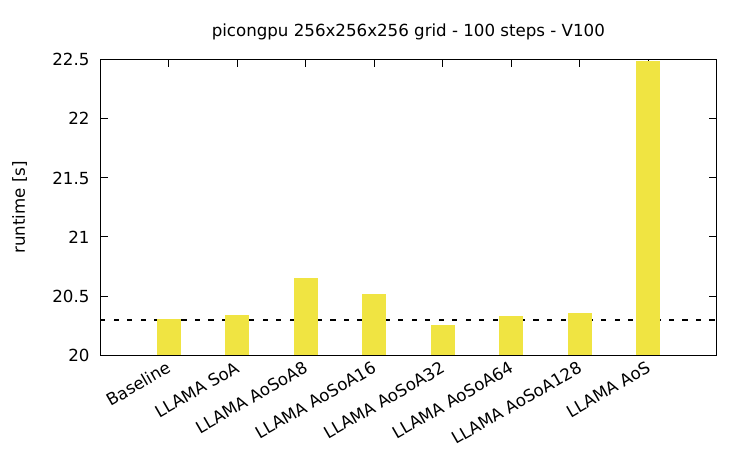}}
			\caption{
				Runtime comparison of PIConGPU runs with the built-in and LLAMA provided data structures.
				We report the computation time for 100 timesteps as printed by the simulation.
				The LLAMA SoA mapping, replicating the data structure used in the original PIConGPU version (Baseline), is almost as fast, while the AoSoA32 is slightly faster than the baseline.
				Other AoSoA layouts and the AoS mapping perform worse.
				\label{fig:picongpu}
			}
	\end{minipage}
\end{figure}

For the scope of this article, only the particle frame list data structure is investigated, which is sketched in figure~\ref{fig:picongpu_frame_list}.
Individual frames are linked with two pointers pointing to the previous and next frame, thus forming a doubly linked list.
Each supercell has a pointer to such a linked list containing its local particles.
Each frame then stores the user configured and additional internal attributes for a fixed number of particles, which is configurable but usually 256 to map well to a thread block on GPUs.
The attributes are stored as SoA with padding between the attribute arrays.
As particles move between supercells during the simulation, individual particles may be moved within a frame or between frames, and entire frames may be allocated or deallocated.

Using the Boost~MPL~\cite{boost_mpl}, PIConGPU puts together a type list of all particle attributes and the frame data structure is generated using template meta programming.
PIConGPU also provides a type representing a reference to a single particle stored distributedly inside a frame.
The frame and particle reference could thus easily be replaced by a LLAMA view and a virtual record, respectively.
We would like to point out that the LLAMA SoA mappings cannot insert padding between the sub arrays yet.
But for the specific simulation ran, each frame held 256 particles, which does not result in any padding needed to guarantee the correct alignment of the particle attribute sub arrays.
This shortcoming of the LLAMA version of PIConGPU initially turned out to be faster, because PIConGPU applied too much padding to improve the memory access patterns on older GPU architectures.
This padding has been reduced now and the memory layout matches the one generated by the LLAMA SoA mapping in the presented benchmark.
Furthermore, we implemented an extension to LLAMA's \lstinline{ArrayDims} class allowing it to hold the array dimensions at compile time, making the wrapping LLAMA mappings and view stateless except for the blobs.
Due to the array dimensions being known at compile time, the blob size can also be computed at compile time and we can use a byte array with compile time extent as blob type.
This finally allows to reinterpret a plain byte array, as returned by PIConGPU's allocator mallocMC\cite{mallocMC}, as a LLAMA view.

For the benchmark we used a development version of picongpu\footnote{
	Git commit: \url{https://github.com/ComputationalRadiationPhysics/picongpu/commit/48fc84b3d1396f65633533600c7e7afdef4c1d45}.
} and the simulation found in picongpu/share/picongpu/benchmarks/SPEC.
We ran it on a single node with \lstinline[language=bash]{mpiexec picongpu -d 1 1 1 -g 256 256 256 -s 100 -p 5 --periodic 1 1 1}.
Figure~\ref{fig:picongpu} shows runtime measurements of a PIConGPU simulation without LLAMA (Baseline) and with LLAMA, running on an NVIDIA V100.
The LLAMA SoA version is 0.2\% slower than the baseline.
While it seems that the LLAMA provided SoA layout could match the native data structure in PIConGPU, further investigation and profiling is necessary.
A first look at the disassembly revealed a substantial difference in the generated instruction stream.
We assume that this is caused by the additional stress that LLAMA's abstractions put on the compiler's optimizer, leading to e.g. different loop unrolling decisions.
We would like to emphasize here that the frame list data structure in PIConGPU is at the core of highly optimized production code which we replaced by LLAMA at close-to-zero overhead.
Since the execution of the code on NVIDIA hardware is organized in warps of 32 threads, the LLAMA AoSoA32 mapping is a natural choice, because it increases locality while preserving coalesced access of the warps.
However, the LLAMA AoSoA32 mapping is just 0.2\% faster than the baseline.
Decreasing or increasing the lane count of the AoSoA to 8, 16, 64 or 128 was 1.7\%, 1.1\%, 0.1\% and 0.3\% slower than the baseline, respectively, confirming that AoSoA32 is the sweet spot.
The LLAMA AoS version is 10.7\% slower than the baseline, which was expected because the SoA data layout allows for a coalesced memory access pattern.
Nevertheless, with LLAMA we now have a tool to check such assumptions by changing a single line of code.

\section{Conclusions}
\label{sec_conclusion}

We have presented the C++ library LLAMA, which provides mechanisms for defining and switching memory layouts of data structures at compile time, without the need to change the programs expressed on them, and enabling efficient copies between different layouts.
LLAMA is an attempt to provide a mitigation to the increasing number of programs becoming memory bound on a growing diversity of hardware architectures.
Efficiency of parallel programs increasingly relies on choosing optimal memory layouts to maximize data throughput, increase locality and facilitate parallel access.
For portable codes, the choice of a memory layout thus needs to be decoupled from the rest of a program, requiring a zero-runtime-overhead abstraction layer, underneath which memory layouts can be freely exchanged.
Previous attempts of memory layout optimization lack generality, expressiveness, flexibility or freedom.

The conceptual and library design of LLAMA has been discussed together with a selection of the core APIs to construct a program.
LLAMA is designed around the concept of a target-hardware-independent data space, an abstract definition of a data structure as a mix of compile and runtime information.
Users express programs based on this data space, achieving full independence of the physical data representation in memory.
Memory layouts on the other side are described by LLAMA mappings, which are concise descriptions of how indices of the data space are mapped to physical memory locations.
Enabled by lazy evaluation, mappings have a holistic view and provide full control of each memory access, giving the user a capable toolbox to craft, adjust and fine-tune memory layouts to perfectly fit any hardware architecture, even ones that did not exist yet when they wrote their programs.
Mapping implementations thus benefit from additional information on target hardware and a user program's memory access pattern.
LLAMA does not yet identify, collect or provide this information, but this is a logical next step in our further investigation.
A representation such as the Memory Access Skeleton proposed by Spindle~\cite{spindle} looks suitable to describe the memory access pattern.
LLAMA mappings are specified in terms of contiguous byte ranges, called blobs, as lightweight memory abstraction, allowing a full decoupling of memory layout and allocation.
This way LLAMA can be combined with any third-party allocator or existing memory ranges, allowing use with static memory segments or in environments without dynamic memory allocations.
Since mappings encapsulate all information required to precisely specify a memory layout, LLAMA is able to provide layout aware copies which adapt and optimize depending on the involved layouts, providing superior throughput compared to element-wise copies.
The LLAMA library is written in standard C++17, tapping into a rich and widely used ecosystem that is available on a broad range of systems.
We achieved almost seamless integration of LLAMA into the C++ language and standard library, allowing for expressive use of e.g. the STL algorithms.

LLAMA's performance is close to native implementations.
We have shown and benchmarked the update and move phase of an all-pairs n-body simulation, the former being compute-bound with good caching, the latter being memory-bound with streaming characteristics.
Compared with manually written data structures, we demonstrated that LLAMA's performance on CPUs is on a par for the AoS and SoA mappings.
AoSoA memory layouts additionally require a different loop structure in the user programs, which LLAMA cannot provide yet.
On the GPU, we showed that LLAMA versions with suitable memory layouts perform at least as fast as a manual reference implementation.
Combined with kernel abstraction libraries such as Alpaka, single source kernels, performance portable across multiple hardware architectures, are thus attainable.
When copying data between different memory layouts, we have shown in a further benchmark that LLAMA's layout-aware copy routines can significantly speed up data transfer and reshuffling of data.
With the \speccpu 2017 lbm benchmark, we showed that by integrating LLAMA to control the central data structure, we could reproduce the original performance by picking an equivalent memory layout.
On top of that, LLAMA's Trace mapping allowed us to design a Split mapping to separate hot from cold data.
By changing a single line of code, we could switch to SoA and various AoSoA layouts, allowing for a rich exploration and giving us a further performance boost.
The integration into PIConGPU's particle frame list data structure has shown that LLAMA can successfully be used in a highly complex physics simulation, a direction which we want to further explore.

LLAMA's design is not yet close to native data structure descriptions in C++.
This is not surprising as we are breaking the C++ paradigm that a data structure definition also defines its memory layout.
We think LLAMA could greatly benefit from upcoming facilities~\cite{layout_attribute, metaprogramming_sutton} proposed for C++ as part of the reflection effort\footnote{
	The C++ reflection effort is formally pursued by SG7 of ISO/IEC JTC1/SC22 WG21 (informally: the C++ standard committee) in meetings open to the public.
	Discussed proposals are available at: \url{http://open-std.org/jtc1/sc22/wg21/docs/papers}.
} and would like to provide feedback and suggestions.

In the future, we will focus on testing LLAMA with more real-world applications and on more hardware platforms.
We would like to further enrich LLAMA's mapping capabilities, while staying within the boundaries of the design requirements described in the introduction.
Finally, although we think that LLAMA itself should only provide a framework to formulate memory mappings producing optimal memory layouts, we also expect that building facilities for explicit vectorization or automatic optimum mapping choice are well within the reach of LLAMA's existing capabilities.

\section*{Acknowledgments}

We would like to thank Axel Naumann for setting up the collaboration between CERN and CASUS,
Jan Stephan for the regular talks on ideas and concepts as well as proof-reading this article,
Verena Gruber for proof-reading and consulting on color, layout and design of the figures,
Jiří Vyskočil for discussing and improving the nbody benchmark in its early state,
Simeon Ehrig for proof-reading this article,
and Ivo Sbalzarini for providing detailed information about OpenFPM.

\subsection*{Author contributions}

The corresponding author Bernhard Manfred Gruber composed this article with its benchmarks and is the current developer of the LLAMA library.
Michael Bussmann pioneered the initial ideas and vision for LLAMA, provided guidance and feedback on this article and continues to provide conceptual input to LLAMA's development.
Rene Widera provided several additional foundational ideas on which LLAMA was built and continues to guide the library's development.
Alexander Matthes was the previous developer of the LLAMA library and laid out the foundation of the C++ library.
Jakob Blomer greatly helped with organizing, revising and proofreading this article.
Guilherme Amadio provided great help with optimizing and tweaking the benchmarks and plots and provided feedback on this article.

\subsection*{Financial disclosure}

This work has been sponsored by the Wolfgang Gentner Programme of the German Federal Ministry of Education and Research (grant no. 05E18CHA).
This project has received funding from the European Union’s Horizon 2020 research and innovation programme under grant agreement No 654220.
This work was partially funded by the Center of Advanced Systems Understanding (CASUS) which is financed by Germany's Federal Ministry of Education and Research (BMBF) and by the Saxon Ministry for Science, Culture and Tourism (SMWK) with tax funds on the basis of the budget approved by the Saxon State Parliament.

\subsection*{Conflict of interest}

The authors declare no potential conflict of interests.

\section*{Supporting information}

LLAMA is developed on GitHub: \url{https://github.com/alpaka-group/llama}, where we provide further documentation and also host the n-body and layout changing copy benchmarks shown in this article.

\appendix

\section{Manual and LLAMA SoA n-body disassembly comparison}
\noindent\begin{minipage}{.48\textwidth}
	\begin{lstlisting}[language={[x86masm]Assembler},basicstyle=\tiny\ttfamily,caption=Manual n-body SoA disassembly (AVX512 version).,label=lst:manual_soa_disassembly]
push    rbp
mov     r11, rcx
mov     r10, r8
vmovaps zmm10, ZMMWORD PTR .LC0[rip]
vmovaps zmm9, ZMMWORD PTR .LC1[rip]
xor     r8d, r8d
vmovaps zmm8, ZMMWORD PTR .LC2[rip]
mov     rbp, rsp
mov     rcx, QWORD PTR [rbp+16]


.L3:
vmovups zmm13, ZMMWORD PTR [rdi+r8]
vmovups zmm12, ZMMWORD PTR [rsi+r8]
xor     eax, eax
vmovups zmm11, ZMMWORD PTR [rdx+r8]
vmovups zmm7, ZMMWORD PTR [r11+r8]
vmovups zmm6, ZMMWORD PTR [r10+r8]
vmovups zmm5, ZMMWORD PTR [r9+r8]
.L2:
vaddps  zmm4, zmm13, DWORD PTR [rdi+rax*4]{1to16}
vaddps  zmm3, zmm12, DWORD PTR [rsi+rax*4]{1to16}
vaddps  zmm2, zmm11, DWORD PTR [rdx+rax*4]{1to16}
vmulps  zmm4, zmm4, zmm4
vmulps  zmm3, zmm3, zmm3
vmulps  zmm2, zmm2, zmm2
vaddps  zmm0, zmm4, zmm10
vaddps  zmm0, zmm0, zmm3
vaddps  zmm1, zmm0, zmm2
vmulps  zmm0, zmm1, zmm1
vmulps  zmm0, zmm0, zmm1
vsqrtps zmm0, zmm0
vdivps  zmm0, zmm9, zmm0
vmulps  zmm0, zmm0, DWORD PTR [rcx+rax*4]<{1to16}>
add     rax, 1
vmulps  zmm0, zmm0, zmm8
vfmadd231ps     zmm7, zmm0, zmm4
vfmadd231ps     zmm6, zmm0, zmm3
vfmadd231ps     zmm5, zmm0, zmm2
cmp     rax, 16384
jne     .L2
vmovups ZMMWORD PTR [r11+r8], zmm7
vmovups ZMMWORD PTR [r10+r8], zmm6
vmovups ZMMWORD PTR [r9+r8], zmm5
add     r8, 64
cmp     r8, 65536
jne     .L3
vzeroupper
pop     rbp
ret
	\end{lstlisting}
\end{minipage}\hfill
\begin{minipage}{.48\textwidth}
	\begin{lstlisting}[language={[x86masm]Assembler},basicstyle=\tiny\ttfamily,caption=LLAMA n-body SoA mapping disassembly (AVX512 version).,label=lst:llama_soa_disassembly]
mov     rsi, QWORD PTR [rdi+8]
mov     rcx, QWORD PTR [rdi+32]
xor     r8d, r8d
mov     rdx, QWORD PTR [rdi+56]
mov     r11, QWORD PTR [rdi+80]
mov     r10, QWORD PTR [rdi+104]
mov     r9, QWORD PTR [rdi+128]
vmovaps zmm10, ZMMWORD PTR .LC1[rip]
mov     rdi, QWORD PTR [rdi+152]
vmovaps zmm9, ZMMWORD PTR .LC2[rip]
vmovaps zmm8, ZMMWORD PTR .LC3[rip]
.L12:
vmovups zmm13, ZMMWORD PTR [rsi+r8]
vmovups zmm12, ZMMWORD PTR [rcx+r8]
xor     eax, eax
vmovups zmm11, ZMMWORD PTR [rdx+r8]
vmovups zmm7, ZMMWORD PTR [r11+r8]
vmovups zmm6, ZMMWORD PTR [r10+r8]
vmovups zmm5, ZMMWORD PTR [r9+r8]
.L11:
vsubps  zmm4, zmm13, DWORD PTR [rsi+rax*4]{1to16}
vsubps  zmm3, zmm12, DWORD PTR [rcx+rax*4]{1to16}
vsubps  zmm2, zmm11, DWORD PTR [rdx+rax*4]{1to16}
vmulps  zmm4, zmm4, zmm4
vmulps  zmm3, zmm3, zmm3
vmulps  zmm2, zmm2, zmm2
vaddps  zmm0, zmm4, zmm10
vaddps  zmm0, zmm0, zmm3
vaddps  zmm1, zmm0, zmm2
vmulps  zmm0, zmm1, zmm1
vmulps  zmm0, zmm0, zmm1
vsqrtps zmm0, zmm0
vdivps  zmm0, zmm9, zmm0
vmulps  zmm0, zmm0, DWORD PTR [rdi+rax*4]{1to16}
add     rax, 1
vmulps  zmm0, zmm0, zmm8
vfmadd231ps     zmm7, zmm0, zmm4
vfmadd231ps     zmm6, zmm0, zmm3
vfmadd231ps     zmm5, zmm0, zmm2
cmp     rax, 16384
jne     .L11
vmovups ZMMWORD PTR [r11+r8], zmm7
vmovups ZMMWORD PTR [r10+r8], zmm6
vmovups ZMMWORD PTR [r9+r8], zmm5
add     r8, 64
cmp     r8, 65536
jne     .L12
vzeroupper

ret
	\end{lstlisting}
\end{minipage}

\bibliography{bibliography}

\clearpage

\section*{Author Biography}

\setlength{\tabcolsep}{0pt}
\newlength\biocolumn
\setlength{\biocolumn}{\textwidth-72pt}

\noindent\begin{tabular}{p{66pt}@{\hskip 6pt}p{\biocolumn}}
	\vspace{0cm}\includegraphics[width=66pt,height=86pt]{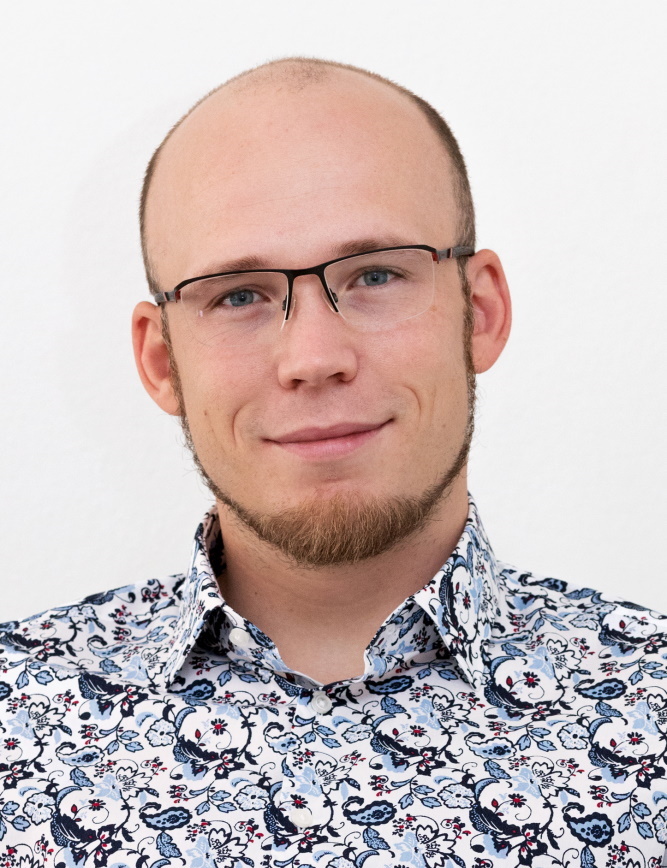} &
	\textbf{Bernhard Manfred Gruber} is a PhD student at CERN in the department of software for experimental physics, where he focuses on performance portable memory layout abstractions in C++.
	He previously worked as a software engineer at CERN on real-time software and embedded PROFINET firmware for the cryogenics instrumentation and quench protection of the LHC and the ITER fusion experiment.
	Before CERN, Bernhard worked at RISC Software GmbH on geometric modeling tools for aerospace engineering and the simulation and visualization of industrial machining processes.
	He lectured briefly at the University of Applied Sciences Upper Austria, from which he received an M.Sc. in Software Engineering in 2016.
\end{tabular}

\vspace{6pt}

\noindent\begin{tabular}{p{66pt}@{\hskip 6pt}p{\biocolumn}}
	\vspace{0cm}\includegraphics[width=66pt,height=86pt]{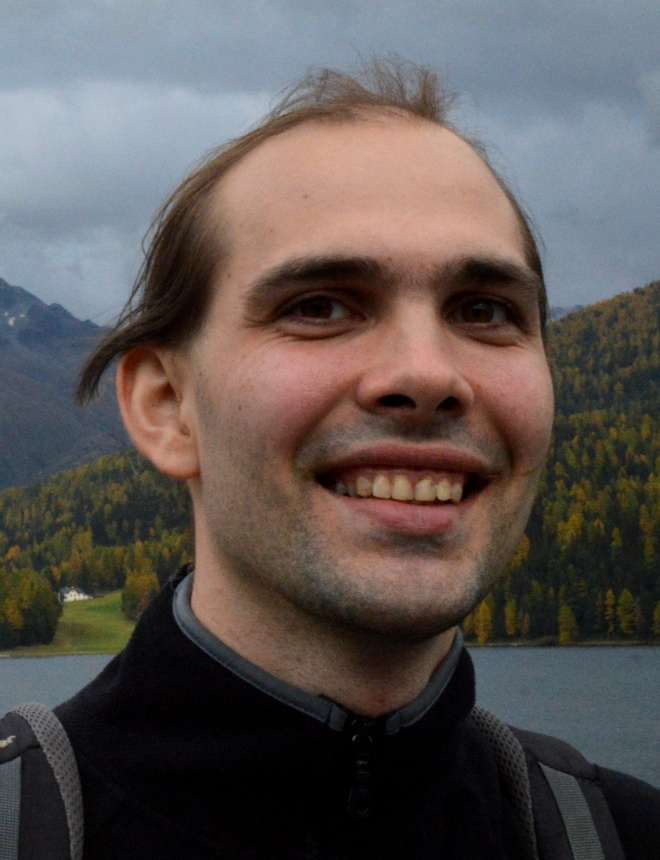} &
	\textbf{Alexander Matthes} studied computer science at the TU Dresden.
	During his studies he worked on developing and maintaining simulations and visualizations of a particle accelerator and a high-performance laser facility at the Helmholtz-Zentrum Dresden - Rossendorf.
	His diploma thesis on in-situ visualization of simulations got him in touch with HPC GPUs, many- and multi-core architectures.
	With the gained expertise in C++ template meta programming, he created an abstraction library for high performance memory access on different architecture.
	Later, Alex developed tools for remote sensing of resources using satellite images.
	In 2020 he joined LogMeIn to work on state-of-the-art video conference solutions.
\end{tabular}

\vspace{6pt}

\noindent\begin{tabular}{p{66pt}@{\hskip 6pt}p{\biocolumn}}
	\vspace{0cm}\includegraphics[width=66pt,height=86pt]{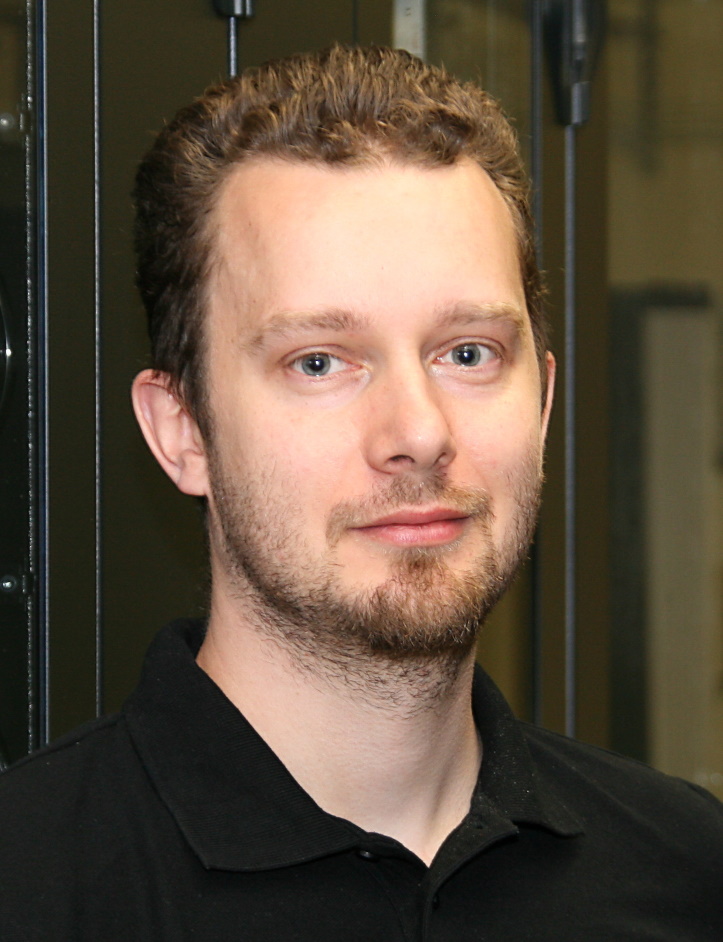} &
	\textbf{René Widera} finished his education as IT specialist for application development at the Technical University Dresden (Germany) in 2009.
	Since 2019 he leads the technical software developemnt parts of the Computational Radiation Physics group at Helmholtz-Zentrum Dresden-Rossendorf (Germany).
	\emph{PIConGPU} the world’s fastest 3D3V electromagnetic particle-in-cell code for plasma physics and the heterogeneous kernel abstraction library \emph{alpaka} are the most prominent applications and libraries he is contributing.
	He is a co-author of “Radiative Signatures of the Relativistic Kelvin-Helmholtz Instability”, one of the six finalists of the Super Computing 2013 Gordon Bell competition.
\end{tabular}

\vspace{6pt}

\noindent\begin{tabular}{p{66pt}@{\hskip 6pt}p{\biocolumn}}
	\vspace{0cm}\includegraphics[width=66pt,height=86pt]{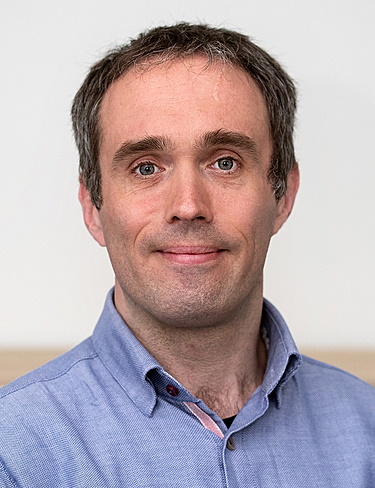} &
	\textbf{Michael Bussmann} is founding manager of the Center for Advanced systems Understanding.
	A theoretical plasma physicist by training, Michael's interest are in high performance computing, real-time, streaming and large-scale data analytics, machine learning, in-situ visualization and portable solutions for parallel programming.
	Michael is active in bridging the gap across disciplines, working in astrophysics, particle physics, accelerator physics, photon science and medical physics.
\end{tabular}

\vspace{6pt}

\noindent\begin{tabular}{p{66pt}@{\hskip 6pt}p{\biocolumn}}
	\vspace{0cm}\includegraphics[width=66pt,height=86pt]{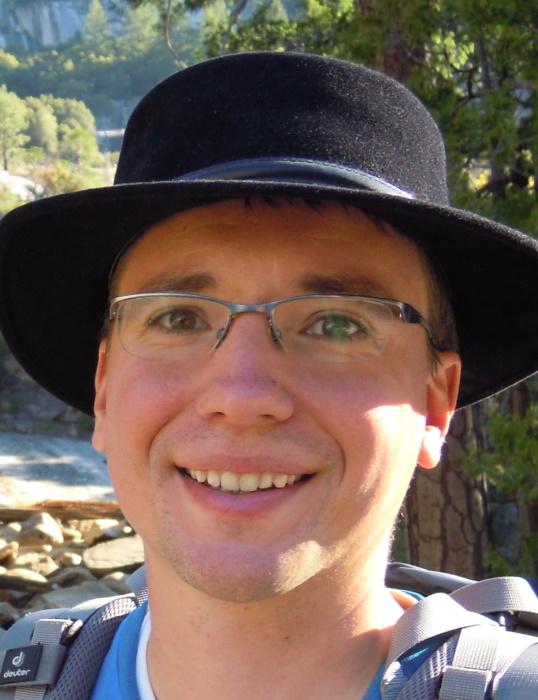} &
	\textbf{Jakob Blomer} is a staff computer scientist in the scientific software group at CERN.
	His research interests include distributed storage systems and data organization for analytics.
	He authored the CernVM File System and he drives the R\&D effort on ROOT RNTuple, a columnar file format for HEP.
	Jakob received a PhD in computer science from the Technical University of Munich in 2012.
	Before his current post, he was a Marie Curie fellow at CERN and a visiting scholar at Stanford University.
\end{tabular}

\vspace{6pt}

\noindent\begin{tabular}{p{66pt}@{\hskip 6pt}p{\biocolumn}}
	\vspace{0cm}\includegraphics[width=66pt,height=86pt]{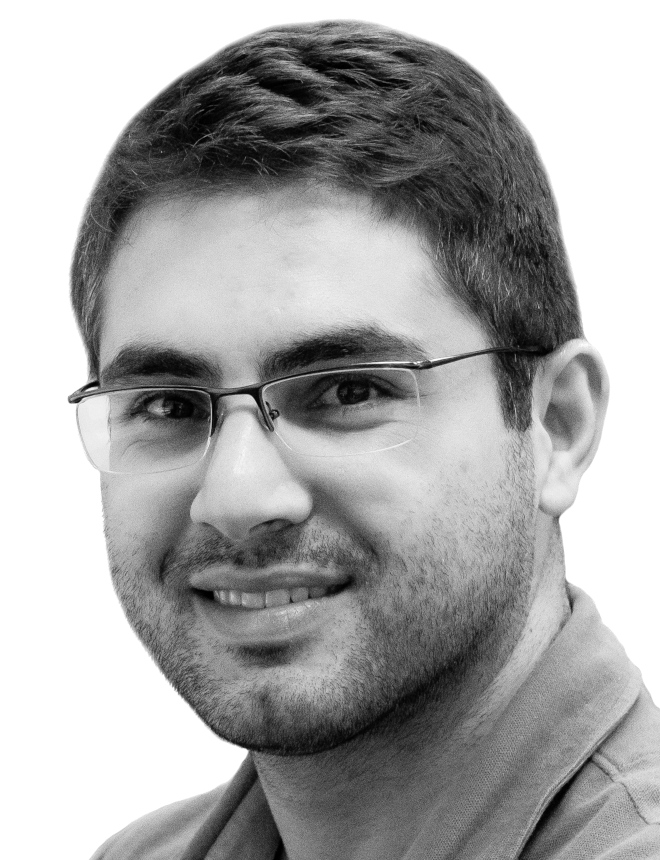} &
	\textbf{Guilherme Amadio} is a scientific software developer at CERN.
	His work focuses on performance optimization of Monte Carlo detector simulations and on research for porting them onto GPUs.
	He received his Ph.D. in aerospace engineering from the University of Illinois at Urbana-Chamapaign in 2014, and M.Sc. degree in physics from the University of Tokyo in 2007.
\end{tabular}

\end{document}